\documentclass[3p,12pt]{elsarticle}
\usepackage{threeparttable,booktabs,tabularx}
\usepackage[fleqn]{amsmath}
\usepackage{cases}
\usepackage{multirow}
\usepackage{xcolor}
\usepackage[normalem]{ulem}
\usepackage{tabularx}
\usepackage{siunitx}
\usepackage{comment}
\usepackage{algorithm}
\usepackage{algpseudocode}
\usepackage{algorithmicx}
\usepackage{grffile}
\usepackage{rotating}
\usepackage{array}
\usepackage{lineno}
\usepackage{hyperref}
\usepackage{makecell}
\usepackage{graphicx,enumerate}
\usepackage{amssymb}
\usepackage{bm,amsmath}
\allowdisplaybreaks  
\usepackage{subfigure}
\usepackage{caption,color}
\usepackage{threeparttable}
\usepackage{subeqnarray}
\usepackage{multirow}
\usepackage{lscape}
\usepackage{rotating}
\usepackage{graphics}
\floatname{algorithm}{Algorithm}
\usepackage{epstopdf}
\journal{}

\newcounter{bla}

\usepackage{CJKutf8}

\begin{document}
\begin{CJK}{UTF8}{gbsn}


\begin{frontmatter}

\title{Adjoint shape optimization of oscillatory rarefied gas flows}

\author{Pengshuo Li} 
\author{Lei Wu\corref{mycorrespondingauthor}}
\cortext[mycorrespondingauthor]{Corresponding author}
\ead{wul@sustech.edu.cn}
 
\address{Department of Mechanics and Aerospace Engineering, Southern University of Science and Technology, Shenzhen 518055, China}

\begin{abstract}
A fast-converging and asymptotic-preserving adjoint shape optimization method is proposed for drag reduction of multiscale gas flows in vibrating micro-electro-mechanical systems. The convergence of the Boltzmann kinetic equation is accelerated by macroscopic synthetic equations, whose constitutive relations integrate continuum-limit terms and high-order kinetic corrections to faithfully characterize spatiotemporal rarefaction effects. As such, this method maintains near-continuum limit consistency while retaining high kinetic accuracy in rarefied flow regimes. Fourier stability analysis performed in an infinite domain demonstrates that the present method yields a spectral radius below 0.5, indicating that the numerical deviation from the converged solution is halved per iteration. Numerical simulations are conducted on an oscillating cylinder and a comb-shaped resonator. The results verify the high accuracy of the derived adjoint sensitivities and the excellent drag reduction performance of the proposed method across various Knudsen and Strouhal numbers. Compared with conventional kinetic iteration methods, the present method produces convergent primal and adjoint solutions within dozens of iterations and features asymptotic preserving behavior, permitting spatial cell sizes far larger than the molecular mean free path. This facilitates efficient design of vibrating micro-electro-mechanical systems.
\end{abstract}


\end{frontmatter}
\section{Introduction}

Micro-electro-mechanical systems (MEMS), including resonant sensors, accelerometers, gyroscopes and comb-drive resonators, find widespread use in sensing, actuation and signal processing \cite{karniadakis2005microflows,tang1989laterally}. During operation, ambient gas exerts damping forces on oscillating micro-structures. This directly governs the quality factor, resonant frequency, bandwidth and power consumption of MEMS devices \cite{bao2007squeeze}.  Consequently, geometry optimization to mitigate gas-induced losses is critical for boosting MEMS performance.

While flow optimization based on the Euler or Navier–Stokes (NS) equations has been extensively investigated \cite{jameson1988aerodynamic,giles2000adjoint}, optimization targeting rarefied gas flows encountered in MEMS necessitates the Boltzmann kinetic equation: at low pressures or high oscillation frequencies, spatial and temporal rarefaction effects become pronounced \cite{li2026frequency}. Under such conditions, the NS equations fail to predict non-equilibrium phenomena reliably. 
The Boltzmann equation governs the velocity distribution function defined within six-dimensional phase space, which imposes severe computational burdens. For this reason, adjoint optimization represents the method of choice, as it computes gradients of the objective function with respect to numerous design variables at a computational cost nearly independent of the number of design parameters. The feasibility of adjoint kinetic optimization has only recently been demonstrated \cite{sato2019topology, caflisch2021adjoint, guan2023topology, yuan2025adjoint,zhang2026adjoint}; nevertheless, efficient adjoint optimization for oscillatory rarefied gas flows remains unexplored.

The primal and adjoint kinetic equations can be solved via stochastic or deterministic methods. The direct simulation Monte Carlo method is a dominant stochastic technique for rarefied gas flow simulations \cite{bird1994molecular}. Nevertheless, it suffers from prominent statistical noise when simulating low-speed MEMS oscillatory flows. Additionally, its inherent time-marching framework leads to extremely high computational costs for low-frequency oscillatory flow simulations.
The variance-reduction~\cite{baker2005variance,homolle2007low} and frequency-domain~\cite{ladiges2015frequency,ladiges2015deviational} Monte Carlo methods resolve the aforementioned noise and computational inefficiency issues, respectively, yet they still pose heavy computational burdens for near-continuum flows, as the time step and spatial cell size must be respectively smaller than the mean collision time and mean free path. As an alternative, deterministic discrete velocity method eliminates sampling noise, rendering it well-suited for simulating small-amplitude oscillatory flows \cite{wu2014oscillatory}. However, like the traditional Monte Carlo methods, conventional iterative schemes for deterministic solvers exhibit slow convergence and large numerical dissipation in the near-continuum regime \cite{wang2018comparative}. From a computational perspective, an efficient multiscale numerical method should ideally possess fast-converging and asymptotic-preserving properties. The former enables steady-state solutions to be achieved within dozens of iterations \cite{su2020steady}, while the latter admits spatial cell sizes substantially larger than the molecular mean free path \cite{Jin2022_APschemes}.

The deterministic general synthetic iterative scheme (GSIS) provides a powerful strategy for accelerating multiscale kinetic simulations. It achieves fast-converging and asymptotic-preserving properties through the simultaneous solution of mesoscopic kinetic and macroscopic synthetic systems \cite{su2020steady,su2020fast}. Specifically, the kinetic system supplies high-order closure terms to constrain the macroscopic synthetic equations. In return, steady-state solutions of these synthetic equations efficiently guide the evolution of the velocity distribution function toward convergence. Thanks to this bidirectional coupling mechanism, the GSIS framework delivers rapid convergence and asymptotic consistency with the NS framework in the near-continuum regime.
A frequency-domain GSIS has recently been developed for linear oscillatory rarefied gas flows \cite{li2026frequency}, which allows complex periodic flow responses to be computed without resolving lengthy transient evolution \cite{wang2022investigation}. These favorable properties motivate the development of an efficient adjoint GSIS framework for shape optimization of oscillating MEMS devices in rarefied gas conditions.

The remainder of this paper is organized as follows. Section~\ref{sec:kinetic_model} presents the frequency-domain linearized kinetic equation, the mesoscopic adjoint formulation, and gas-kinetic boundary conditions. Section~\ref{sec:GSIS} derives the constitutive relations for the adjoint macroscopic equation in near-continuum  regime, develops the GSIS scheme for the adjoint kinetic equation, and compares the convergence rates of the conventional iterative scheme and the GSIS scheme. Section~\ref{sec:shape_sensitivity} elaborates the discrete shape sensitivity analysis and the optimization framework based on the free-form deformation. Sections~\ref{sec:cylinder_optimization} and~\ref{sec:ba} validate the proposed method through shape optimization cases of an oscillating cylinder and a biaxial accelerometer, respectively. Finally, conclusions are drawn in Section~\ref{sec:conclusion}.

\section{Formulation}
\label{sec:kinetic_model}

In this section, we formulate the adjoint shape optimization problem for linear oscillatory rarefied gas flows. We first introduce the frequency-domain linearized kinetic equation and the gas--wall interaction models. The mesoscopic adjoint formulation is then derived for the prescribed objective functional, followed by a summary of the overall optimization procedure.


\subsection{Frequency-domain linearized kinetic equation}
\label{subsec:forward_kinetic_equation}

We consider a general MEMS structure undergoing small-amplitude harmonic oscillation. The velocity of the moving boundary \(\Gamma_m\) is prescribed as
\begin{equation}\label{wall_velocity}
    \bm{U}_w(t)=\Re\{U_0\bm{e}_m\exp(\mathrm{i}\omega t)\},
\end{equation}
where $U_0$ and $\omega$ denote the vibration amplitude and angular frequency, \(\bm{e}_m\) is the unit vector showing the vibration direction, $\mathrm{i}$ is the imaginary unit, $t$ is the time, and $\Re\{\cdot\}$ denotes the real part of a complex number. The vibration amplitude is assumed to be sufficiently small compared with the most probable molecular velocity $v_m$, namely,
\begin{equation}\label{normalized}
    \xi=\frac{U_0}{v_m}\ll 1, \quad \text{with}~ v_m= \sqrt{\frac{2k_B T_0}{m}},
\end{equation}
where $k_B$ is the Boltzmann constant and $m$ is the molecular mass of the gas.
Under this assumption, the gas response can be described by a linearized kinetic model in the frequency domain~\cite{ladiges2015frequency,ladiges2015deviational}.

For a small-amplitude harmonic oscillation satisfying $\xi\ll1$, once the periodic state is established, the velocity distribution function in the Boltzmann kinetic equation is expressed as
\begin{equation}
    f(\boldsymbol{v},\boldsymbol{x},t)
    =
    \Re{
    \left\{
    f_{\mathrm{eq}}(\boldsymbol{v})
    \left[
    1+\xi h(\boldsymbol{v},\boldsymbol{x})
    \exp(i\omega t)
    \right]
    \right\}
    }
,
\end{equation}
where $\bm{v}=({v}_x,{v}_y,{v}_z)$ is the molecular velocity normalized by $v_m$, $\bm{x}=(x,y,z)$ is the spatial coordinate vector normalized by the reference length $L_\text{ref}$, and $h$ denotes the perturbation from the global equilibrium state $f_{eq}(\bm{v})=\pi^{-3/2}\exp(-|\bm v|^2)$. Let $\Omega\subset\mathbb{R}^{3}$ be the physical domain and $\Xi\subset\mathbb{R}^{3}$ be the molecular velocity space. The frequency-domain linearized Shakhov kinetic equation considered in this work reads~\cite{shakhov1968approximate}
\begin{equation}
    iS h+\boldsymbol{v}\cdot\nabla h
    =
    \delta_{rp}
    (\mathcal{L}h-h),
    \quad (\bm x,\bm v)\in\Omega\times\Xi,
    \label{eq:linearized_kinetic}
\end{equation}
where $S$ is the Strouhal number and $\delta_{\mathrm{rp}}$ is the rarefaction parameter defined as
\begin{equation}
    S=\frac{\omega L_{\text{ref}}}{v_m},
    \qquad
    \delta_{rp}=\frac{p_0L_{\text{ref}}}{\mu v_m},
\end{equation}
where $p_0$ is the reference pressure, and $\mu$ is the dynamic viscosity of the gas at the reference temperature $T_0$. Note that the rarefaction parameter is related to the spatial Knudsen number $Kn$ (which is defined as the ratio of the mean free path of gas molecules to the reference characteristic length) as $\delta_{rp}=\sqrt{\pi}/(2Kn)$. The temporal Knudsen number $Kn_t$, which is defined as the ratio of the mean collision time of gas molecules to the reference characteristic time, can be expressed as \cite{wu2022rarefied}
\begin{equation}\label{temporal_Kn}
    Kn_t=\delta^{-1}_{rp} S.
\end{equation}

For any velocity-dependent quantity $a(\bm v)$, we use the notation $ \langle a\rangle = \int_{\Xi} a(\bm v) f_{eq}(\bm v)\,d\Xi$ to describe the macroscopic quantities. For examples, the macroscopic quantities, such as perturbation density $\rho$, flow velocity $\bm u$, perturbation temperature $\tau$, and heat flux $\bm q$, which are respectively normalized by \(\xi\rho_0=\xi mp_0/k_BT_0\), \(\xi v_m\), \(\xi T_0\), \(\xi p_0\), and \(\xi p_0v_m\), 
are
\begin{equation}\label{primal_variable}
\begin{aligned}
    \bm{W} = [\rho,\,\bm u,\,\tau,\,\bm q]^\top
    = \left\langle \bm\chi(\bm v)h \right\rangle,
    \quad \text{with} \quad 
    \bm{\chi}(\bm v)=
   \left[1, \bm v, \dfrac23 |\bm v|^2-1, 
    \left(|\bm v|^2-\dfrac52\right)\bm v\right]^\top,
\end{aligned}
\end{equation}
where $\top$ represents the transpose of a matrix.

The gain part of the linearized collision term reads
\begin{equation}
    \mathcal{L}h = \bm\psi(\bm v)\cdot \bm W,
     \quad \text{with} \quad 
    \bm{\psi}(\bm v)=
    \left[1, 2\bm v, |\bm v|^2-\dfrac32, \dfrac{4}{15}\left(|\bm v|^2-\dfrac52\right)\bm v \right]^\top .
\end{equation}

It should be noted that, the dimensional density $\widetilde{\rho}$, velocity $\widetilde{\bm{U}}$, temperature $\widetilde{T}$, stress tensor $\widetilde{\bm{P}}$ and heat flux $\widetilde{\bm{q}}$, are expressed as
\begin{equation}
\begin{aligned}
    \widetilde{\rho} &= \rho_0\left[1+\xi\Re\left\{\rho(\bm{x})\exp(\mathrm{i}\omega t)\right\}\right],\\
    \widetilde{\bm{U}} &= \xi v_m\Re\left\{\bm{u}(\bm{x})\exp(\mathrm{i}\omega t)\right\},\\
    \widetilde{T} &= T_0\left[1+\xi\Re\left\{\tau(\bm{x})\exp(\mathrm{i}\omega t)\right\}\right],\\
    \widetilde{\bm{P}} &= 
    \widetilde{\rho}R\widetilde{T}
    \left[\bm{I}+\xi\Re\left\{\bm{\Pi}(\bm{x})\exp(\mathrm{i}\omega t)\right\}\right],\\
    \widetilde{\bm{q}} &= \xi p_0v_m\Re\left\{\bm{q}(\bm{x})\exp(\mathrm{i}\omega t)\right\},
\end{aligned}
\label{eq:dimensional_macroscopic_fields}
\end{equation}
where the normalized deviatoric stress tensor is
\begin{equation}
    {\Pi}_{ij}=2\left\langle \left(v_i v_j-\frac13|\bm v|^2\delta_{ij}\right) h \right\rangle.
\end{equation}




\subsection{Boundary condition}
\label{subsec:forward_boundary_condition}

The gas-wall interaction plays an essential role in determining the force
response acting on the vibrating structure. In this work, diffuse reflection
is adopted for all solid boundaries. For the MEMS configurations considered
here, the wall boundary is decomposed as
\[
\Gamma_w=\Gamma_m\cup\Gamma_s ,
\]
where \(\Gamma_m\) denotes the moving wall boundary and \(\Gamma_s\)
denotes the stationary wall boundary. Let $\bm{n}$ denote the unit normal vector pointing from the gas domain toward the wall. The incoming and outgoing velocity spaces are defined as
\begin{equation}
    \Xi^+=\{\boldsymbol{v}:\boldsymbol{v}\cdot\boldsymbol{n}>0\},
    \qquad
    \Xi^-=\{\boldsymbol{v}:\boldsymbol{v}\cdot\boldsymbol{n}<0\}.
\end{equation}

According to the diffuse reflection boundary condition, the outgoing distribution function can be expressed as
\begin{equation}
    h(\bm x,\bm v)
    = \underbrace{2\sqrt{\pi}
    \int_{\Xi^+}
    (\bm v'\cdot\bm n) h(\bm x,\bm v')
    f_{eq}(\bm v')\,d\Xi'}_{\rho_{r}(\bm x)}
    +\underbrace{2\bm u_w\cdot\bm v-\sqrt{\pi}\,\bm u_w\cdot\bm n}_{h_w(\bm v)} 
    \quad \text{in} \quad \Gamma_w \times\Xi^{-},
    \label{eq:forward_diffuse_bc}
\end{equation}
where the normalized wall velocity is \(\bm{u}_w=\bm{e}_m\) for moving wall boundary \(\Gamma_m\) and zero for stationary wall \(\Gamma_s\). The term \(\rho_r(\bm x)\) is the wall re-emission coefficient determined by the zero-mass-flux condition, while the term \(h_w(\bm v)\) represents the perturbation induced by the wall motion.

At the Dirichlet boundary $\Gamma_d$, the distribution for molecules entering the computational domain is prescribed as
\begin{equation}
    h=h_d
    \qquad \text{on} \qquad
    \Gamma_d\times\Xi^-.
\end{equation}
For the equilibrium far-field condition considered in this work, $h_d=0$.

\subsection{Mesoscopic adjoint formulation}


The optimization aims to minimize the amplitude of the gas force exerted on the moving boundary $\Gamma_m$ in the prescribed vibration direction. Once the periodic state is established, the corresponding dimensional force is written as
\begin{equation}
    {F}(t)
    =
    F_{\text{ref}}
    \Re\left\{
    J\exp(\mathrm{i}\omega t)
    \right\},
    \qquad
    F_{\text{ref}}
    =
    \xi p_0 A_{\mathrm{ref}},
    \label{eq:dimensional_force_response}
\end{equation}
where $J$ is the dimensionless complex force amplitude and $A_{\mathrm{ref}}$ is the reference surface area. For a two-dimensional configuration, $A_{\text{ref}}=L_{\text{ref}}$ and $F$ represents the force per unit out-of-plane depth, whereas $A_{\text{ref}}=L_{\text{ref}}^2$ for a three-dimensional configuration.

The complex force amplitude is evaluated from the molecular momentum flux on the moving boundary \(\Gamma_m\):
\begin{equation}
    J= \int_{\Gamma_m} \int_{\Xi}
    v_n m(\boldsymbol{v})
    h(\boldsymbol{x},\boldsymbol{v})
    f_{\mathrm{eq}}(\boldsymbol{v})
    d\Xi\,d\Gamma ,
    \quad v_n=\bm v\cdot\bm n,
    \label{eq:J}
\end{equation}
where \(m(\bm{v})=2\bm{v}\cdot\bm{e}_m\) is the moment kernel associated with the prescribed velocity direction. Taking the wall velocity as the phase reference, the real part of \({J}\) is the in-phase, damping-related component, whereas its imaginary part is the stiffness-related component. 


To eliminate the dependence of the complex force amplitude derivative on the variation of the primal solution, the following Lagrangian is introduced:
\begin{equation}
    L = J+ I + B_w + B_d,
\end{equation}
where
\begin{equation}
\begin{aligned}
    I &= \int_\Omega\int_\Xi
    \phi\left[
    \mathrm{i}S\,h + \bm v\cdot\nabla h - \delta_{rp}(\mathcal{L}h-h)
    \right]f_{eq}(\bm v)\,d\Xi\,d\Omega ,\\
    B_w &= \int_{\Gamma_w}\int_{\Xi^-}
    \phi_w\left[
    h - \rho_r(\bm x) - h_w(\bm v)
    \right]f_{eq}(\bm v)\,d\Xi\,d\Gamma,\\
    B_d &= \int_{\Gamma_d}\int_{\Xi^-}
    \phi_d\left[
    h - h_d
    \right]f_{eq}(\bm v)\,d\Xi\,d\Gamma,
\end{aligned}
\label{eq:lagrangian_terms}
\end{equation}
with $\phi$, $\phi_w$ and \(\phi_d\) being the Lagrangian multipliers to satisfy the kinetic equation and boundary conditions, respectively. They are both defined in the physical and velocity spaces. By requiring the first variation of the Lagrangian with respect to the primal variable to vanish, the adjoint kinetic equation is obtained: 
\begin{equation}\label{eq:adjoint_shakhov}
\left.
\begin{aligned}
    \mathrm{i}S\, \phi - \bm v \cdot \nabla \phi = \delta_{rp}(\hat{\mathcal{L}} \phi- \phi) ,\quad
    &\mathrm{in}\ \Omega \times \Xi, \\
    \phi_w = -v_n(\phi + m) ,
    \quad &\mathrm{in}\ \Gamma_w \times \Xi^{-}, \\
    \phi = -m + 2\sqrt{\pi}\int_{\Xi^{-}} \phi_w f_{eq}(\bm{v}')d\Xi, 
    \quad &\mathrm{in}\ \Gamma_w \times \Xi^{+}, \\
    \phi = 0, 
    \quad &\mathrm{in}\ \Gamma_d \times \Xi^{+},
\end{aligned}
\right\}
\end{equation}
where the adjoint kinetic equation and macroscopic quantities are defined as
\begin{equation}\label{eq:adjoint_m_c}
\begin{aligned}
    \hat{\mathcal L} \phi =& \bm\chi(\bm v)\cdot \hat{\bm W},\\
    \hat{\bm W} =& [\hat\rho,\,\hat{\bm u},\,\hat\tau,\,\hat{\bm q}]^\top
    = \left\langle \bm\psi(\bm v)\phi \right\rangle.
    \end{aligned}
\end{equation}
The normalization of these adjoint macroscopic  quantities are the same as the corresponding primal variables. 

Once the adjoint variable \(\phi_w\) is solved, the total sensitivity of \(J\) with respect to shape variations \(\Gamma_m\) is obtained from the explicit geometric dependence in the Lagrangian:
\begin{equation}
    dL(\Gamma_m;\delta \Gamma_m) = dJ(\Gamma_m;\delta \Gamma_m) + dB_w(\Gamma_m;\delta \Gamma_m).
\end{equation}

\subsection{Overall framework of the optimization}
\label{subsec:optimization_procedure}

\begin{figure}[t]
    \centering
    \includegraphics[width=0.9\linewidth]{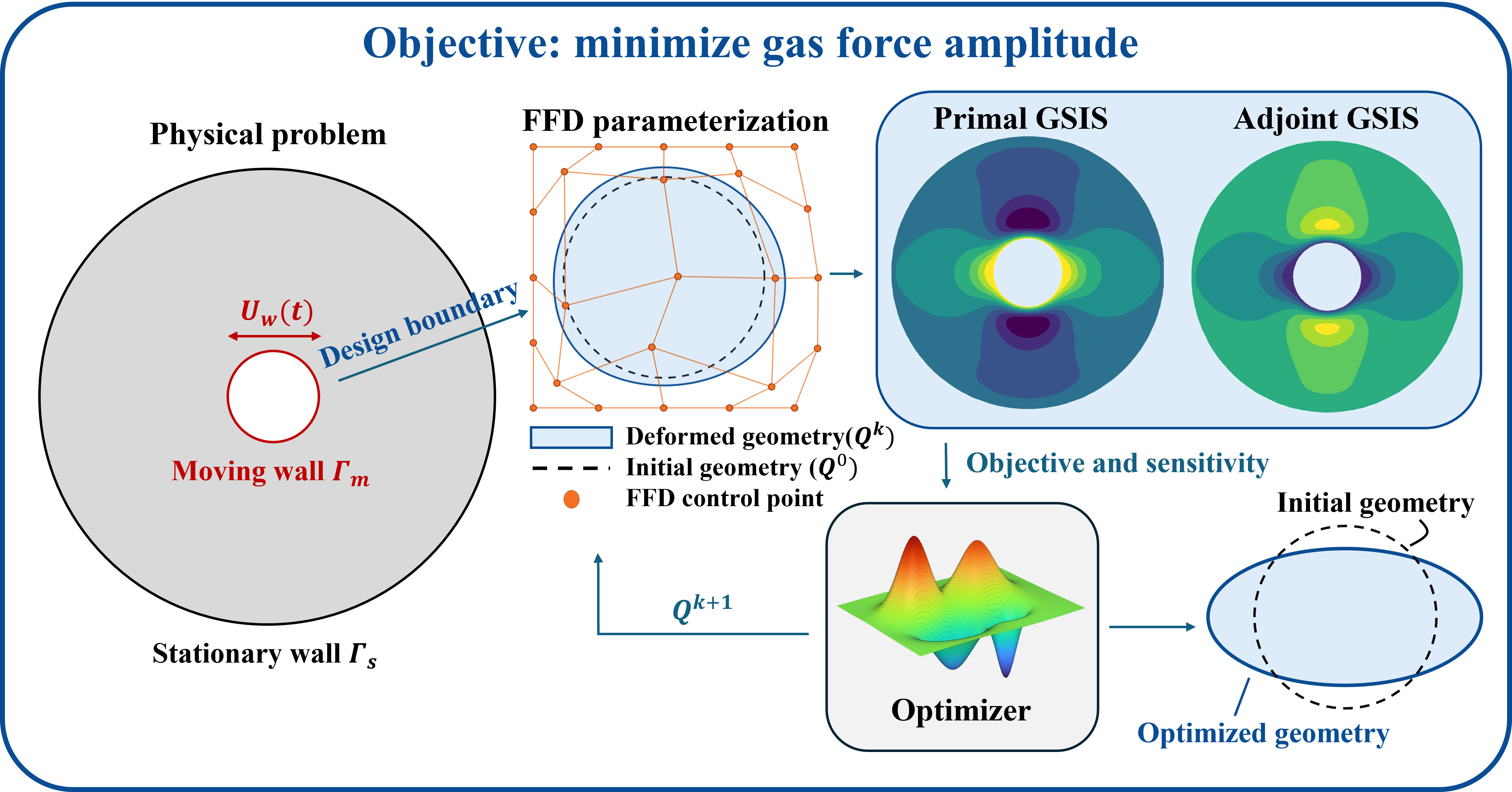}
    \caption{Adjoint optimization framework for periodically‑oscillating gas‑kinetic systems.}
    \label{fig:adjoint-framework}
\end{figure}

The moving boundary $\Gamma_m$ is parameterized by free-form deformation (FFD)~\cite{sederberg1986freeform,samareh2001survey}, and the coordinates of the FFD control points are collected in the design vector $\bm Q$. Since the force response $J$ is complex-valued in the frequency domain, a real-valued objective function is constructed according to the specific optimization target. For the force-amplitude minimization considered in this work, we use $J_{\mathrm{obj}}=|J|^2$, whose sensitivity with respect to the design variables is
\begin{equation}
    \frac{\partial J_{\mathrm{obj}}}{\partial\bm Q}=2\,\Re\left\{J^*\frac{\partial J}{\partial\bm Q}\right\}.
\end{equation}
where $(\cdot)^*$ denotes the complex conjugate. The proposed adjoint shape optimization then follows a gradient-based constrained procedure. As illustrated in Fig.~\ref{fig:adjoint-framework}, starting from $\bm Q^0$, the optimization proceeds as follows:
\begin{enumerate}
    \item Initialize $\bm Q^0$, construct the corresponding moving boundary $\Gamma_m(\bm Q^0)$, and generate the computational mesh.
    \item At the $k$-th iteration, solve the primal and adjoint kinetic equations using GSIS to evaluate $J(\bm Q^k)$ and its sensitivity $\partial J/\partial\bm Q$.
    \item Update the design vector from $\bm Q^k$ to $\bm Q^{k+1}$ using the method of moving asymptotes (MMA)~\cite{svanberg1987method,svanberg2002class}, implemented in NLopt~\cite{NLopt}.
    \item Construct $\Gamma_m(\bm Q^{k+1})$ and deform the interior mesh using the spring smoothing method~\cite{batina1990unsteady}. The procedure terminates when the convergence criterion is satisfied or the maximum number of iterations is reached; otherwise, return to Step~2.
\end{enumerate}

Building upon the frequency-domain GSIS developed in our previous work~\cite{li2026frequency}, the present study establishes an adjoint shape optimization framework for oscillatory rarefied gas flows. A GSIS is further developed for the mesoscopic adjoint equation to ensure efficient solution. The adjoint GSIS and shape-sensitivity formulation are presented in Sections~\ref{sec:GSIS} and~\ref{sec:shape_sensitivity}, respectively.

\section{Adjoint GSIS and asymptotic analysis}\label{sec:GSIS}

 In this section, the GSIS is extended to the adjoint kinetic equation~\eqref{eq:adjoint_shakhov}. We first analyze the convergence behavior of the conventional iterative scheme (CIS), then construct the adjoint synthetic equations and establish the fast-converging and asymptotic-preserving properties of the resulting GSIS. 

\subsection{CIS and its convergence rate}
\label{subsec:adjoint_cis_convergence}

Normally, the adjoint kinetic equation can be straightforwardly solved by the following CIS, which
updates the distribution function by
\begin{equation}
    (\mathrm{i} S+\delta_{rp})\phi^{n+1}
    -
    \bm v\cdot\nabla\phi^{n+1}
    =
    \delta_{rp}\,
    \bm\chi(\bm v)\cdot\hat{\bm W}^{\,n},
    \label{eq:adjoint_cis_iteration}
\end{equation}
where $n$ is the iteration step. That is, given the adjoint distribution function $\phi^n$, the adjoint macroscopic quantities can be calculated as per Eq.~\eqref{eq:adjoint_m_c}. Then, a new distribution at the $(n+1)$-th iteration step is obtained by solving  Eq.~\eqref{eq:adjoint_cis_iteration}. This process is repeated until convergence. 

We use the Fourier stability analysis to calculate the convergence speed of the CIS. Define the error between two consecutive iterations as
\begin{equation}    \label{eq:adjoint_cis_fourier_mode}
\begin{aligned}
    Y^{n+1}
    =&
    \phi^{n+1}-\phi^n
    = e^n\bar Y(\bm v)\exp(\mathrm{i}\bm\theta\cdot\bm x),
    \\
    \bm\Phi^{n+1}
    =&
    \hat{\bm W}^{\,n+1}-\hat{\bm W}^{\,n}
    =
    \left\langle
    \bm\psi Y^{n+1}
    \right\rangle 
=e^{n+1}\bm\alpha\exp(\mathrm{i}\bm\theta\cdot\bm x),
\end{aligned}
\end{equation}
where $\bm{\theta}$ is the perturbation wave vector, $\bm{\alpha}$ is the amplitude vector of the macroscopic error mode,
\(\bm\alpha = [ \alpha_{\hat\rho}, \alpha_{\hat{\bm u}}, \alpha_{\hat\tau}, \alpha_{\hat{\bm q}}]^\top\), and $e$ is the spectral radius.
Without loss of generality, we set \(|\bm{\theta}|=1\) to focus on the dependence on the spatial Knudsen number and Strouhal number.

Substituting Eqs.~\eqref{eq:adjoint_cis_fourier_mode} into Eq.~\eqref{eq:adjoint_cis_iteration} yields
\begin{equation}
    e\bm\alpha
    =
    \left\langle
    \bm\psi(\bm v)\bar Y(\bm v)
    \right\rangle
    ,\qquad
    \bar Y(\bm v)
    =
    \frac{
    \bm\chi(\bm v)\cdot\bm\alpha
    }{
    1+\mathrm{i}\delta_{rp}^{-1}
    \left(
    S-\bm\theta\cdot\bm v
    \right)
    } .
    \label{eq:adjoint_cis_Ybar}
\end{equation}
or equivalently
\begin{equation}
     e\bm\alpha = \bm C_{\rm CIS}^{adj}\,\bm\alpha,\qquad
    \bm C_{\rm CIS}^{adj}
    =
    \left\langle
    \frac{
    \bm\psi(\bm v)\bm\chi^\top(\bm v)
    }{
    1+\mathrm{i}\delta_{rp}^{-1}
    \left(
    S-\bm\theta\cdot\bm v
    \right)
    }
    \right\rangle .
    \label{eq:adjoint_cis_eigen_problem}
\end{equation}
Therefore, the convergence rate of CIS is determined by $\rho(\bm C_{\rm CIS}^{adj})=\max_j|\lambda_j|$,
where $\lambda_j$ is the eigenvalue of the matrix $\bm C_{\rm CIS}^{adj}$.

\begin{figure}[t]
    \centering
    \includegraphics[width=0.55\linewidth]{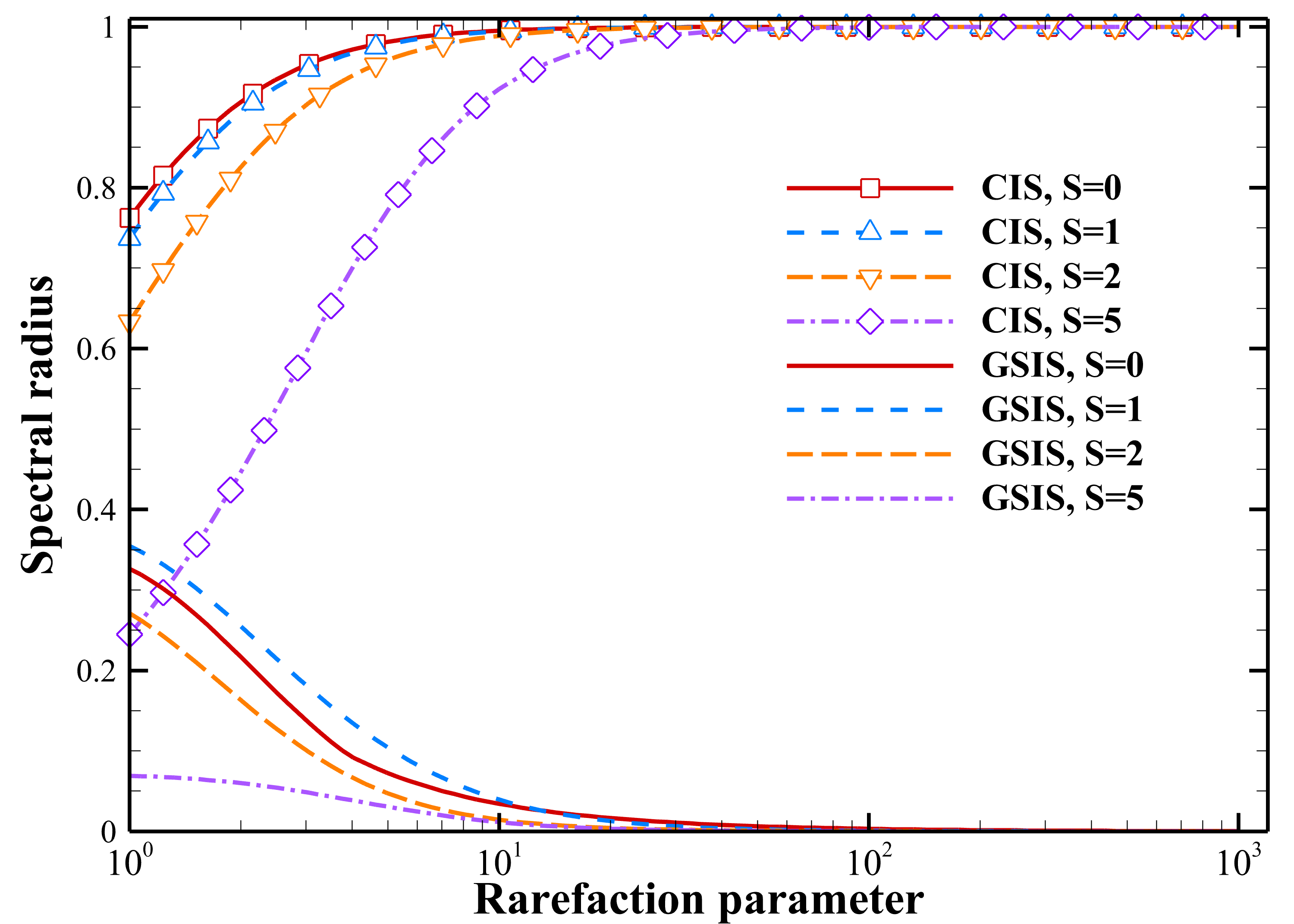}
    \caption{Spectral radius of CIS and GSIS for the adjoint kinetic equation, as functions of the rarefaction parameter \(\delta_{rp}\) for different Strouhal number \(S\).}
    \label{fig:spectral radius}
\end{figure}

It is interesting to note that the convergence matrix of the CIS for the forward equation can be derived as follows~\cite{li2026frequency}:
\begin{equation}
    \bm C_{\rm CIS}^{fwd}
    =
    \left\langle
    \frac{
    \bm\chi(\bm v)\bm\psi^\top(\bm v)
    }{
    1+\mathrm{i}\delta_{rp}^{-1}
    \left(
    S+\bm\theta\cdot\bm v
    \right)
    }
    \right\rangle = \bm C_{\rm CIS}^{adj}(-\bm\theta)^\top.
\end{equation}
Therefore, the primal and adjoint CIS have the same spectral radius.
Hence, the adjoint CIS inherits the same convergence behavior as the primal CIS: as shown in Fig.~\ref{fig:spectral radius}, it is fast in rarefied regimes, while in the near-continuum regime, the spectral radius approaches unity, leading to slow convergence.

\subsection{Adjoint general synthetic iterative scheme}
\label{sec:adjoint_gsis}

To boost the convergence, the adjoint general synthetic iterative scheme (GSIS) is constructed by coupling the adjoint kinetic equation with a set of macroscopic synthetic equations. These equations are obtained from the velocity moments of the adjoint equation \eqref{eq:adjoint_shakhov}, while their continuum closure is derived from the Chapman--Enskog expansion \cite{chapman1990mathematical}.  

Throughout this section, we define \(\partial_i\equiv \partial/\partial x_i\), and the
Einstein summation convention is used.  Multiplying the adjoint kinetic equation by
$2$, $2v_i$, and $|\bm v|^2-3/2$, respectively, and integrating over velocity
space, we have 
\begin{equation}
\begin{aligned}
    &2\mathrm{i}S\hat\rho-\partial_i\hat u_i=0,\\
    &\mathrm{i}S\hat u_i
    -\partial_i\left(\hat\rho+\frac23\hat\tau\right)
    -\partial_j\hat\Pi_{ij}=0,\\
    &\mathrm{i}S\hat\tau
    -\frac12\partial_i\hat u_i
    -\frac{15}{4}\partial_i\hat q_i=0,
\end{aligned}
\label{eq:adjoint_macro_moment_equations}
\end{equation}
where 
\begin{equation}
    \hat\Pi_{ij}
    = 2\left\langle
    \left(v_i v_j-\frac13|\bm v|^2\delta_{ij}\right)\phi
    \right\rangle
    \label{eq:adjoint_stress_definition}
\end{equation}
is the adjoint non-equilibrium stress tensor. These moment equations are exact yet unclosed, as the adjoint stress tensor \(\hat{\Pi}_{ij}\) and heat flux \(\hat{q}_i\) remain high-order kinetic moments that cannot be fully expressed using only low-order adjoint macroscopic quantities.

Inspired by the GSIS~\cite{su2020steady}, the evolution equations for non-equilibrium stress and heat flux are obtained by multiplying Eq.~\eqref{eq:adjoint_shakhov} by $2\left(v_i v_j-\frac13|\bm v|^2\delta_{ij}\right)$ and $\frac{4}{15}\left(|\bm v|^2-\frac52\right)v_i$, respectively.  This gives
\begin{equation}
\begin{aligned}
    &\mathrm{i}S\hat\Pi_{ij}
    -\partial_k \hat M_{ijk}
    = -\delta_{rp}\hat\Pi_{ij},\\
    &\mathrm{i}S\hat q_i
    -\partial_j \hat Q_{ij}
    = -\frac23\delta_{rp}\hat q_i,
\end{aligned}
\label{eq:adjoint_stress_heat_evolution}
\end{equation}
where the higher-order moments are
\begin{equation}\label{eq:adjoint_M_moment}
\begin{aligned}
    \hat M_{ijk}
    &=2\left\langle
    \left(v_i v_j-\frac13|\bm v|^2\delta_{ij}\right)v_k\phi
    \right\rangle, \\
    \hat Q_{ij}
    &=\frac{4}{15}\left\langle
    \left(|\bm v|^2-\frac52\right)v_i v_j\phi
    \right\rangle .
\end{aligned}
\end{equation}

The core idea of the GSIS is to explicitly introduce continuum constitutive relations into Eq.~\eqref{eq:adjoint_M_moment}. Specifically, the stress and heat flux are decomposed into NS constitutive relations and high-order terms (HoTs) that capture rarefaction effects:
\begin{equation}    \label{eq:adjoint_nsf_hot_decomposition}
\begin{aligned}
    \hat\Pi_{ij}
    &=
    \hat\Pi_{ij}^{\text{NS}}
    +
    \hat\Pi_{ij}^{\text{HoT}},
    \\
    \hat q_i
    &=
    \hat q_i^{\text{NS}}
    +
    \hat q_i^{\text{HoT}},
\end{aligned}
\end{equation}
where the HoTs of stress and heat flux are defined as
\begin{equation}
\begin{aligned}
    \hat\Pi_{ij}^{\text{HoT}} &=
    \frac{1}{\mathrm{i}S+\delta_{rp}}\partial_k\hat M_{ijk}
    - \hat\Pi_{ij}^{\text{NS}},\\
    \hat q_i^{\text{HoT}} &=
    \frac{1}{\mathrm{i}S+\frac23\delta_{rp}}\partial_j\hat Q_{ij}
    - \hat q_i^{\text{NS}}.
\end{aligned}
\label{eq:adjoint_hot_definition}
\end{equation}
The NS components are expressed using velocity and temperature gradients, whereas the HoTs are derived from the velocity distribution function and can only be accurately evaluated via numerical solutions of the kinetic equation. This decomposition enables rapid convergence: while the hyperbolic kinetic equation only affects regions within a few molecular mean free paths, the diffusive NS equations enable stable, efficient propagation of flow information across the entire computational domain. 

Now we use the Chapman--Enskog expansion to derive the continuum constitutive relations for the adjoint kinetic equation.
In the near-continuum regime, introduce \(\varepsilon=\delta_{rp}^{-1}\), assume that \(Kn,Kn_t\ll1\), such that \(\varepsilon,\varepsilon S\ll1\).  
The adjoint kinetic equation becomes
$\hat{\mathcal L}\phi-\phi
    =\varepsilon\left(\mathrm{i}S\phi-\bm v\cdot\nabla\phi\right)$.
The adjoint distribution function is expanded in the form $\phi=\phi^{(0)}+\varepsilon\phi^{(1)}+O(\varepsilon^2)$, and macroscopic quantities are expanded following the identical form. Thus, collecting the $O(\epsilon^0)$ term, we have the equilibrium distribution:
\begin{equation}
    \phi^{(0)}
    =\hat\rho+\bm v\cdot\hat{\bm u}
    +\left(\frac23|\bm v|^2-1\right)\hat\tau .
    \label{eq:adjoint_zeroth_order_distribution}
\end{equation}
Consequently, $ \hat{\bm q}^{(0)}=\bm 0$ and $\hat{\bm\Pi}^{(0)}=\bm 0$. Collecting the $O(\epsilon)$ term, we have
$\hat{\mathcal L}\phi^{(1)}-\phi^{(1)}
    =\mathrm{i}S\phi^{(0)}-\bm v\cdot\nabla\phi^{(0)}$.
Taking the stress and heat-flux moments, we get the adjoint NS closure:
\begin{equation}
\begin{aligned}
    \label{eq:adjoint_nsf_closure}
    &\hat\Pi_{ij}^{\text{NS}}
    =\frac{1}{2\delta_{rp}}\left(
    \partial_i\hat u_j+\partial_j\hat u_i
    -\frac23\delta_{ij}\partial_k\hat u_k
    \right),
    \\
    &\hat q_i^{\text{NS}}
    =\frac{1}{3\delta_{rp}}\partial_i\hat\tau .    
\end{aligned}
\end{equation}
It is noted that the adjoint NS relations have the same tensorial structure as their primal counterparts, but the signs follow from the adjoint transport operator and coefficient is different from the primal one~\cite{li2026frequency}. This NS constitutive relation is only approximate, with the truncation error \(O(\epsilon^2)\).


Substituting Eqs.~\eqref{eq:adjoint_nsf_hot_decomposition} into \eqref{eq:adjoint_macro_moment_equations} gives the adjoint synthetic
macroscopic equations
\begin{equation}
\begin{aligned}
    &2\mathrm{i}S\hat\rho-\partial_i\hat u_i=0,\\
    &\mathrm{i}S\hat u_i
    -\partial_i\left(\hat\rho+\frac23\hat\tau\right)
    -\partial_j\hat\Pi_{ij}^{\text{NS}}
    =\partial_j\hat\Pi_{ij}^{\text{HoT}},\\
    &\mathrm{i}S\hat\tau
    -\frac12\partial_i\hat u_i
    -\frac{15}{4}\partial_i\hat q_i^{\text{NS}}
    =\frac{15}{4}\partial_i\hat q_i^{\text{HoT}}.
\end{aligned}
\label{eq:adjoint_synthetic_macro_equations_rho}
\end{equation}
The left-hand side contains the continuum adjoint NS operator, while the right-hand side contains the kinetic correction evaluated from the adjoint distribution function. 

\subsection{Convergence rate analysis}
\label{subsec:adjoint_gsis_convergence}

We now analyze the convergence rate of the adjoint GSIS. For simplicity, the spatial derivative is kept intact; that of the discretized kinetic equation will be shown in numerical simulations. At the \(n\)-th iteration, the intermediate adjoint distribution is obtained by one CIS update:
\begin{equation}
    (\mathrm{i}S+\delta_{rp})\phi^{n+\frac12}
    -
    \bm v\cdot\nabla\phi^{n+\frac12}
    =
    \delta_{rp}\,
    \bm\chi(\bm v)\cdot\hat{\bm W}^{\,n}.
    \label{eq:adjoint_gsis_half_step}
\end{equation}
The high-order terms \(\hat\Pi_{ij}^{HoT,n+\frac12}\) and \(\hat q_i^{HoT,n+\frac12}\) are then evaluated from \(\phi^{n+\frac12}\) according to Eq.~\eqref{eq:adjoint_hot_definition}, and the updated macroscopic variables are obtained by solving the synthetic equations.

Define the kinetic half-step error and the macroscopic error as
\begin{equation} \label{eq:adjoint_gsis_fourier_mode}
\begin{aligned}
    Y^{n+\frac12}
    =&
    \phi^{n+\frac12}-\phi^n
    =e^n\bar Y(\bm v)\exp(\mathrm{i}\bm\theta\cdot\bm x),
    \\
    \bm\Phi^{n+1}
    =&
    \hat{\bm W}^{\,n+1}-\hat{\bm W}^{\,n}
    =e^{n+1}\bm\alpha\exp(\mathrm{i}\bm\theta\cdot\bm x).
\end{aligned}
\end{equation}
It turns out that $\bar Y(\bm v)$ is identical to its counterpart in CIS, while the spectral radius $e$ is determined by the following linear systems:
\begin{equation}
\begin{aligned}
    &e\left(
    2(\mathrm{i}S\alpha_{\hat\rho}
    - \mathrm{i}\theta_i\alpha_{\hat u_i}
    \right) =0,\\
    &e\left[
    \mathrm{i}S\alpha_{\hat u_i}
    - \mathrm{i}\theta_i
    \left( \alpha_{\hat\rho} + \frac23\alpha_{\hat\tau} \right)
    + \frac{1}{2\delta_{rp}}
    \left( \alpha_{\hat u_i}
    + \frac13\theta_i\theta_j\alpha_{\hat u_j} \right) \right]
    = \mathcal S_i^{\Pi},\\
    &e\left(
    \mathrm{i}S\alpha_{\hat\tau}
    - \frac12 \mathrm{i}\theta_i\alpha_{\hat u_i}
    - \frac{15}{4}\mathrm{i}\theta_i\alpha_{\hat q_i}
    \right) =0, \\
    &e\left(
    \alpha_{\hat q_i} -
    \frac{\mathrm{i}\theta_i}{3\delta_{rp}}\alpha_{\hat\tau}
    \right) = \mathcal S_i^{q},
\end{aligned}
\label{eq:adjoint_gsis_fourier_system}
\end{equation}
where \(i=1,2,3\), and the source terms come from the HoTs evaluated at the kinetic half step:
\begin{equation}
\begin{aligned}
    &\mathcal S_i^{\Pi}
    = \left\langle
    \left[ \frac{
    v_i + \frac13\theta_i v_\theta}{\delta_{rp}}
    - \frac{2v_\theta}{\mathrm{i}S+\delta_{rp}}
    \left(v_i v_\theta - \frac13 v^2\theta_i \right) \right]
    \bar Y \right\rangle ,\\
    &\mathcal S_i^{q}
    = \left\langle \left[
    - \frac{\mathrm{i}\theta_i}{3\delta_{rp}}
    \left( v^2-\frac32 \right) +
    \frac{4\mathrm{i}}{ 15\left(\mathrm{i}S+\frac23\delta_{rp}\right)}
    \left( v^2-\frac52 \right)
    v_i v_\theta \right]
    \bar Y \right\rangle,
\end{aligned}
\end{equation}
with \(v_\theta =\bm v\cdot \bm\theta\).

The above system can be eventually written compactly as $ e\bm L\bm\alpha = \bm R\bm\alpha$. The error amplification matrix of the adjoint GSIS is then $\bm G = \bm L^{-1}\bm R$, and the convergence rate $ \rho(\bm G)$ is characterized by its eigenvalue of the largest magnitude. The numerical results are plotted in Fig.~\ref{fig:spectral radius}.
In the near-continuum regime, the spectral radius scales as
\begin{equation}
    \rho(\bm G)
    =
    O(\delta_{rp}^{-2}),
    \qquad
    \delta_{rp}\rightarrow\infty ,
    \label{eq:adjoint_gsis_super_convergence_scaling}
\end{equation}
so that the false convergence of CIS is transformed into the super-convergence
of GSIS:
\begin{equation}
    \frac{\rho(\bm G)}{1-\rho(\bm G)}\epsilon
    \rightarrow
    \frac{\epsilon}{\delta_{rp}^{2}},
    \qquad
    \delta_{rp}\rightarrow\infty .
    \label{eq:adjoint_gsis_false_convergence_removed}
\end{equation}
Thus the adjoint GSIS remains rapidly convergent in the near-continuum,
low-frequency regime, while the CIS spectral radius approaches unity.

\subsection{Overview of the adjoint GSIS}
\label{subsec:adjoint_gsis_overview}

The adjoint GSIS procedure is summarized as follows:
\begin{enumerate}
    \item Given \(\phi^n\) and \(\widehat{\bm U}^{\,n}=[\hat{\rho}^n,\hat{\bm{u}}^n ,\hat{\tau}^n]^\top\), solve Eq.~\eqref{eq:adjoint_gsis_half_step} to obtain the intermediate distribution \(\phi^{n+\frac{1}{2}}\). In this paper, the finite-volume method is adopted, and the detailed numerical scheme is given in \ref{sec:finite_v_kinetic}.

    \item Evaluate the macroscopic moments and the high-order terms $\widehat{\bm\Pi}^{\mathrm{HoT},n+\frac{1}{2}}$ and $\widehat{\bm q}^{\mathrm{HoT},n+\frac{1}{2}}$ from $\phi^{n+\frac{1}{2}}$.
    
    \item Solve the macroscopic synthetic equation \eqref{eq:adjoint_synthetic_macro_equations_rho}, with the numerical method in~\ref{sec:finite_v_gsis}.

    \item Correct the intermediate distribution using the updated macroscopic variables:
    \begin{equation}
        \phi^{n+1} =
        \phi^{n+\frac{1}{2}}
        + \left( 
        \hat{\rho}^{n+1} - \hat{\rho}^{n+\frac{1}{2}}
        \right)
        + \bm v\cdot
        \left(
        \hat{\bm u}^{n+1} - \hat{\bm u}^{n+\frac{1}{2}}
        \right)
        + \left(\frac{2}{3}|\bm v|^2-1 \right)
        \left(
        \hat{\tau}^{n+1} - \hat{\tau}^{n+\frac{1}{2}}
        \right),
        \label{eq:adjoint_macro_micro_correction}
    \end{equation}
\end{enumerate}
so that the density, velocity, and temperature of the velocity distribution function are guided by the solution of macroscopic synthetic equation.  
These steps are repeated until the prescribed convergence criterion is satisfied.







\section{Adjoint sensitivity analysis}
\label{sec:shape_sensitivity}

In this section, the shape derivative is evaluated using the converged primal and adjoint solutions. The discrete boundary contributions to the Lagrangian are first differentiated with respect to the face geometry. The resulting face sensitivities are then transferred to the boundary nodes and projected onto the FFD design variables through the chain rule.

\subsection{Discrete boundary functional and adjoint sensitivity evaluation}
\label{subsec:discrete_boundary_functional}

In the finite-volume method, the deformable boundary $\Gamma_m$ is discretized into faces indexed by $l$. The centroid and outward unit normal vector of the $l$-th face are denoted by $\bm{x}_l$ and $\bm{n}_l$, respectively, while $A_l$ denotes its edge length in two dimensions or face area in three dimensions. The outgoing and incoming discrete velocity sets at the $l$-th face are defined as $\Xi_l^{+}=\left\{k:\bm{v}_k\cdot\bm{n}_l>0\right\}$ and $\Xi_l^{-}=\left\{k:\bm{v}_k\cdot\bm{n}_l<0\right\}$, respectively.

Following Eq.~\eqref{eq:J}, the discrete force-response functional on $\Gamma_m$ is written as
\begin{equation}
    J=\sum_{l\in\Gamma_m}A_l
    \left[
    \sum_{k=1}^{N_{\text{vel}}}
    (\bm v_k\cdot\bm n_l)
    m(\bm v_k)
    h_{l,k}
    f_{eq}(\bm v_k)
    \omega_k
    \right],
    \label{eq:discrete_J}
\end{equation}
where $N_{\text{vel}}$ is the total number of discrete velocities and $\omega_k$ is the corresponding quadrature weight. The moving-wall contribution to the Lagrangian is discretized as
\begin{equation}
    B_m=
    \sum_{l\in\Gamma_m}
    A_l
    \left\{
    \sum_{k\in\Xi_l^-}
    \phi_{w,l,k}
    \left[
    h_{l,k}
    -
    \rho_{r,l}
    -
    h_{w,l,k}
    \right]
    f_{eq}(\bm v_k)
    \omega_k
    \right\},
    \label{eq:discrete_Bm}
\end{equation}
where $\rho_{r,l}
    =
    2\sqrt{\pi}
    \sum_{k\in\Xi_l^+}
    (\bm v_k\cdot\bm n_l)
    h_{l,k}
    f_{eq}(\bm v_k)
    \omega_k$ and $h_{w,l,k}
    =
    2\bm u_w\cdot\bm v_k
    -
    \sqrt{\pi}\,\bm u_w\cdot\bm n_l$.

After the primal and adjoint equations have converged, the dependence of the objective derivative on the variation of the primal solution is eliminated by the adjoint formulation. The remaining shape derivative is therefore evaluated from the explicit dependence of Eqs.~\eqref{eq:discrete_J} and \eqref{eq:discrete_Bm} on $A_l$, $\bm x_l$, and $\bm n_l$.

\subsection{Sensitivities with respect to face geometry}
\label{subsec:face_geometry_derivatives}

Treating $A_l$, $\bm x_l$, and $\bm n_l$ as temporarily independent geometric variables, the partial derivatives of $J$ are
\begin{equation}\label{eq:dJ} 
\begin{aligned}
    \frac{\partial J}{\partial A_l}
    =&\sum_{k=1}^{N_{\text{vel}}}
    (\bm v_k\cdot\bm n_l)m(\bm v_k) h_{l,k}f_{eq}(\bm v_k)\omega_k,\\
    \frac{\partial J}{\partial \bm x_l}
    =&A_l\sum_{k=1}^{N_{\text{vel}}}
    (\bm v_k\cdot\bm n_l)m(\bm v_k)
    \frac{\partial h_{l,k}}{\partial\bm x_l}
    f_{eq}(\bm v_k)\omega_k,\\
    \frac{\partial J}{\partial \bm n_l}
    =&A_l\sum_{k=1}^{N_{\text{vel}}}
    m(\bm v_k) h_{l,k}\bm v_k f_{eq}(\bm v_k)\omega_k .
\end{aligned}
\end{equation}
where the spatial derivative $\partial h_{l,k}/\partial\bm x_l$ is evaluated using the Gauss formula~\cite{yuan2025adjoint,zhang2026adjoint}.

Consequently, the partial derivatives of $B_m$ are
\begin{equation} \label{eq:dBm}
\begin{aligned}
    \frac{\partial B_m}{\partial A_l}&=0,\\
    \frac{\partial B_m}{\partial \bm x_l}
    &=A_l\sum_{k\in\Xi_l^-}
    \phi_{w,l,k}
    \left(
    \frac{\partial h_{l,k}}{\partial\bm x_l}
    -\frac{\partial \rho_{r,l}}{\partial\bm x_l}
    -\frac{\partial h_{w,l,k}}{\partial\bm x_l}
    \right)
    f_{eq}(\bm v_k)\omega_k,\\
    \frac{\partial B_m}{\partial \bm n_l}
    &=A_l\sum_{k\in\Xi_l^-}
    \phi_{w,l,k}
    \left(
    -\frac{\partial \rho_{r,l}}{\partial\bm n_l}
    -\frac{\partial h_{w,l,k}}{\partial\bm n_l}
    \right)
    f_{eq}(\bm v_k)\omega_k .
\end{aligned}
\end{equation}

The derivatives of the wall re-emission coefficient are
\begin{equation}\label{eq:drhor}
\begin{aligned}
    \frac{\partial \rho_{r,l}}{\partial\bm x_l}
    =&2\sqrt{\pi}\sum_{k\in\Xi_l^+}
    (\bm v_k\cdot\bm n_l)
    \frac{\partial h_{l,k}}{\partial\bm x_l}
    f_{eq}(\bm v_k)\omega_k,\\
    \frac{\partial \rho_{r,l}}{\partial\bm n_l}
    =&2\sqrt{\pi}\sum_{k\in\Xi_l^+}
    \bm v_k h_{l,k} f_{eq}(\bm v_k)\omega_k.
\end{aligned}    
\end{equation}

For the prescribed wall velocity considered here, $\bm u_w$ is independent of the local face position and normal direction. Therefore,
\begin{equation}
    \frac{\partial h_{w,l,k}}{\partial\bm x_l}=\bm 0,
    \qquad
    \frac{\partial h_{w,l,k}}{\partial\bm n_l}=-\sqrt{\pi}\,\bm u_w.
    \label{eq:wall_motion_derivatives}
\end{equation}

For a two-dimensional boundary edge connecting $\bm r_1=(x_1,y_1)^\top$ and $\bm r_2=(x_2,y_2)^\top$, the endpoints are ordered such that the normal defined below points outward from the computational domain. Define
\begin{equation}
    \Delta x=x_2-x_1,
    \qquad
    \Delta y=y_2-y_1,
    \qquad
    A=\sqrt{\Delta x^2+\Delta y^2},
    \qquad
    \bm n=\frac{1}{A}
    \begin{bmatrix}
        \Delta y\\
        -\Delta x
    \end{bmatrix}.
    \label{eq:edge_geometry_definition}
\end{equation}
The derivatives of the edge length and centroid with respect to its endpoints are
\begin{equation}
    \frac{\partial A}{\partial\bm r_1}
    =
    \begin{bmatrix}
        -\Delta x/A\\
        -\Delta y/A
    \end{bmatrix},
    \qquad
    \frac{\partial A}{\partial\bm r_2}
    =
    \begin{bmatrix}
        \Delta x/A\\
        \Delta y/A
    \end{bmatrix},
    \qquad
    \frac{\partial\bm x}{\partial\bm r_1}
    =
    \frac{1}{2}\bm I,
    \qquad
    \frac{\partial\bm x}{\partial\bm r_2}
    =
    \frac{1}{2}\bm I,
    \label{eq:edge_area_derivative}
\end{equation}
where $\bm I$ denotes the $2\times2$ identity matrix. The derivatives of the unit normal vector are
\begin{equation}
    \frac{\partial\bm n}{\partial\bm r_1}
    =
    \frac{1}{A^3}
    \begin{bmatrix}
        \Delta x\Delta y & -\Delta x^2\\
        \Delta y^2 & -\Delta x\Delta y
    \end{bmatrix},
    \qquad
    \frac{\partial\bm n}{\partial\bm r_2}
    =
    -
    \frac{\partial\bm n}{\partial\bm r_1}.
    \label{eq:edge_normal_derivative}
\end{equation}

Therefore, for variations of $\Gamma_m$, the sensitivity with respect to a boundary node $\bm r_s$ is obtained by summing the corresponding contributions from $J$ and $B_m$ over all boundary faces connected to that node:
\begin{equation}
    \frac{\partial L}{\partial\bm r_s}
    =
    \sum_{l\in N(s)}
    \left[
    \frac{\partial L}{\partial A_l}
    \frac{\partial A_l}{\partial\bm r_s}
    +
    \left(
    \frac{\partial\bm x_l}{\partial\bm r_s}
    \right)^\top
    \frac{\partial L}{\partial\bm x_l}
    +
    \left(
    \frac{\partial\bm n_l}{\partial\bm r_s}
    \right)^\top
    \frac{\partial L}{\partial\bm n_l}
    \right],
    \label{eq:node_sensitivity}
\end{equation}
where $N(s)$ denotes the set of boundary faces connected to node $s$.

\subsection{Free-form deformation parameterization}
\label{sec:ffd}

Free-form deformation is adopted to parameterize the deformable boundary. By embedding the geometry in a control lattice, FFD generates smooth shape variations through the movement of a limited number of control points~\cite{sederberg1986freeform,samareh2001survey}. Compared with treating all boundary nodes as independent design variables, FFD reduces the dimension of the design space and provides an explicit differentiable mapping from the control-point coordinates to the boundary coordinates.

A two-dimensional FFD lattice is introduced to enclose the deformable boundary. For a boundary node $\bm r_s$, its parametric coordinates $(u_s,v_s)$ are determined from the initial FFD configuration and remain fixed during the deformation. The deformed position of the boundary node is expressed as
\begin{equation}
    \bm r_s(\bm Q)
    =
    \sum_{i=0}^{N_u-1}
    \sum_{j=0}^{N_v-1}
    B_{i,d_u}(u_s)
    B_{j,d_v}(v_s)
    \bm Q_{ij},
    \label{eq:ffd_mapping}
\end{equation}
where $\bm Q_{ij}$ is the coordinate vector of the $(i,j)$-th control point, $N_u$ and $N_v$ are the numbers of control points in the two parametric directions, and $B_{i,d_u}$ and $B_{j,d_v}$ are B-spline basis functions of degrees $d_u$ and $d_v$, respectively~\cite{piegl1997nurbs}.

The geometric derivative of a boundary node with respect to a control point follows directly from Eq.~\eqref{eq:ffd_mapping}:
\begin{equation}
    \frac{\partial\bm r_s}{\partial\bm Q_{ij}}
    =
    B_{i,d_u}(u_s)
    B_{j,d_v}(v_s)
    \bm I.
    \label{eq:ffd_basis_derivative}
\end{equation}

The coordinates of the FFD control points are collected in the
design vector $\bm{Q}$. After each update of $\bm{Q}$, the design boundary $\Gamma_m$ is reconstructed using Eq.~\eqref{eq:ffd_mapping}, and the interior computational mesh is deformed consistently with the updated boundary.

Using the boundary-node sensitivity in Eq.~\eqref{eq:node_sensitivity}, the sensitivity of the Lagrangian with respect to an FFD control point is obtained through the chain rule:
\begin{equation}
    \frac{\partial L}{\partial\bm Q_{ij}}
    =
    \sum_{s\in\Gamma_m}
    \left(
    \frac{\partial\bm r_s}{\partial\bm Q_{ij}}
    \right)^\top
    \frac{\partial L}{\partial\bm r_s}.
    \label{eq:ffd_chain_rule}
\end{equation}

\begin{figure}[p]
    \centering
    \subfigure[computational configuration]{\includegraphics[height=0.35\linewidth]{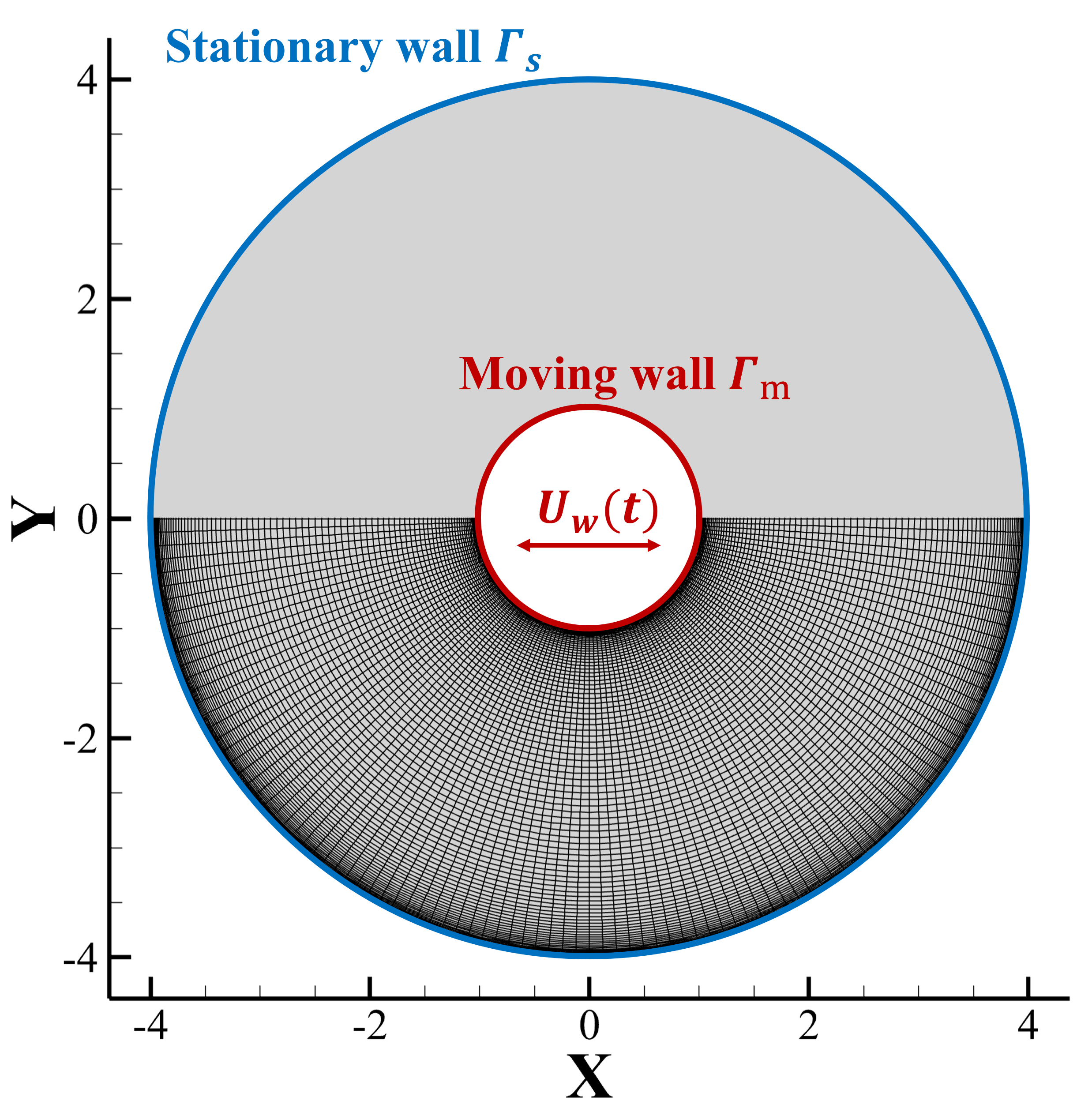}}\\
    \hspace{0.1cm}
    \subfigure[\(\Re(\hat{u}_x)\) when \((\delta_{rp},S)=(100,1)\)]{\includegraphics[height=0.35\linewidth]{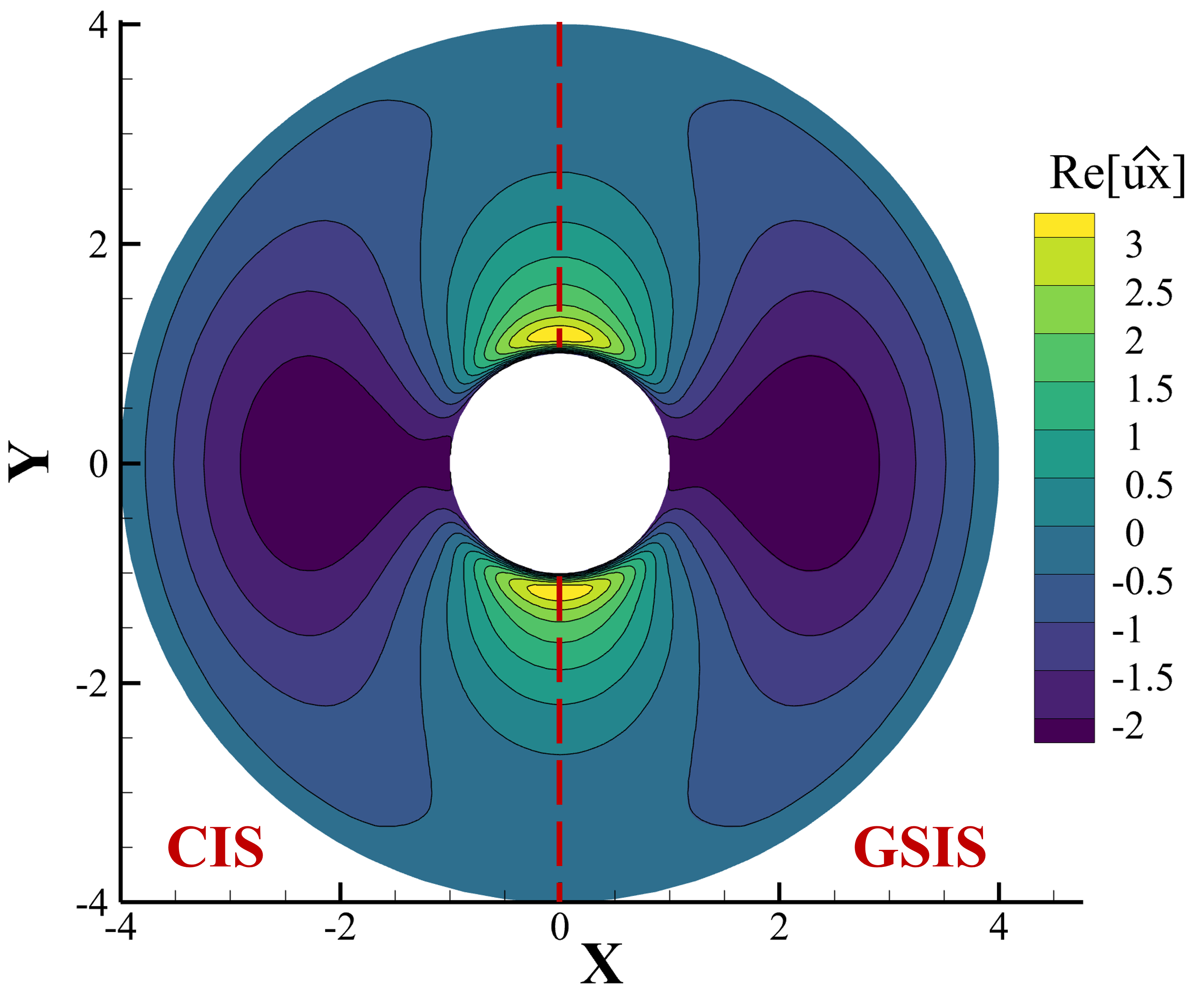}}
    \subfigure[\(\Re(\hat{u}_y)\) when \((\delta_{rp},S)=(100,1)\)]{\includegraphics[height=0.35\linewidth]{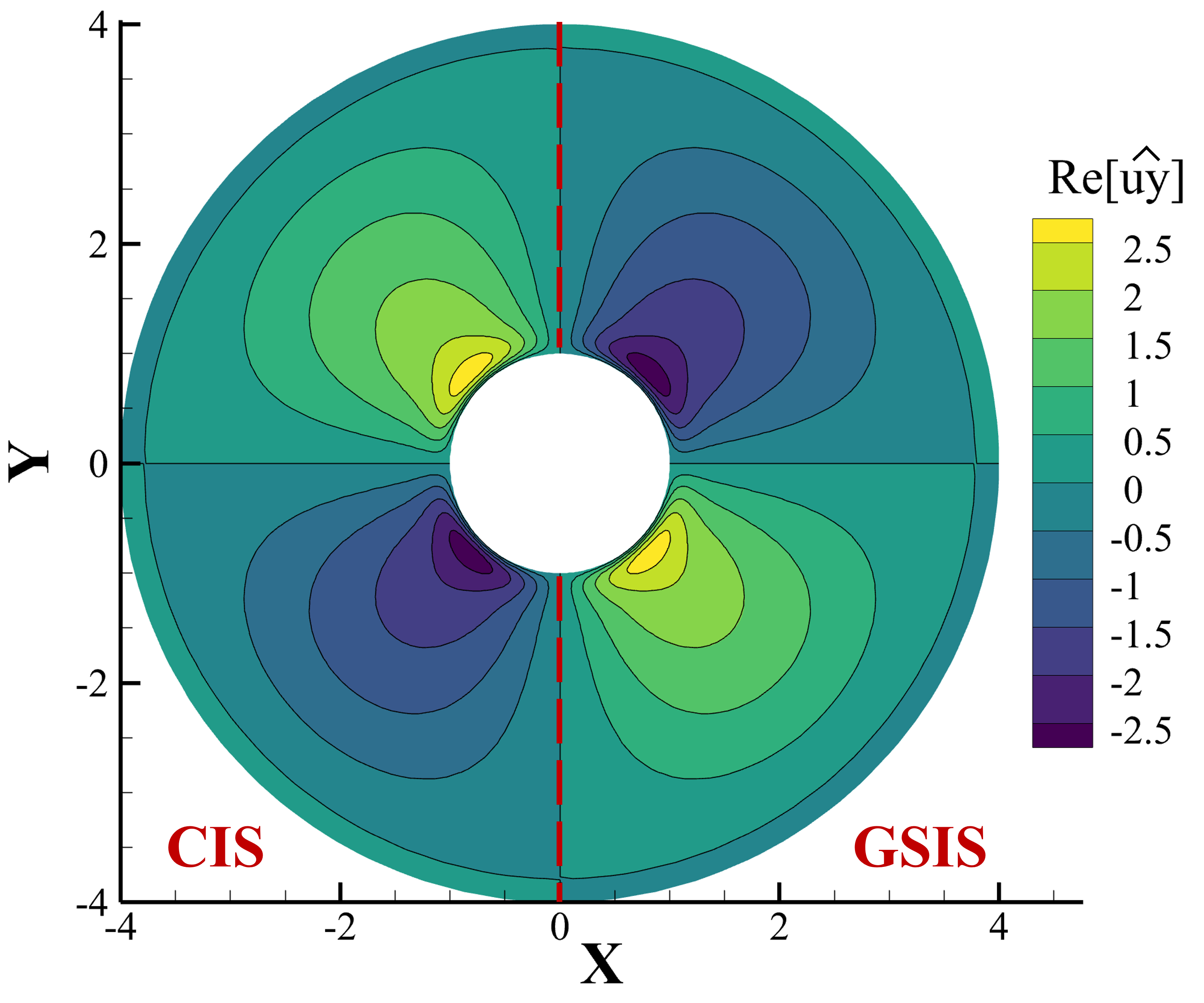}}\\
    \subfigure[\(\Re(\hat{u}_x)\) when \((\delta_{rp},S)=(1000,10^{-5})\)]{\includegraphics[height=0.35\linewidth]{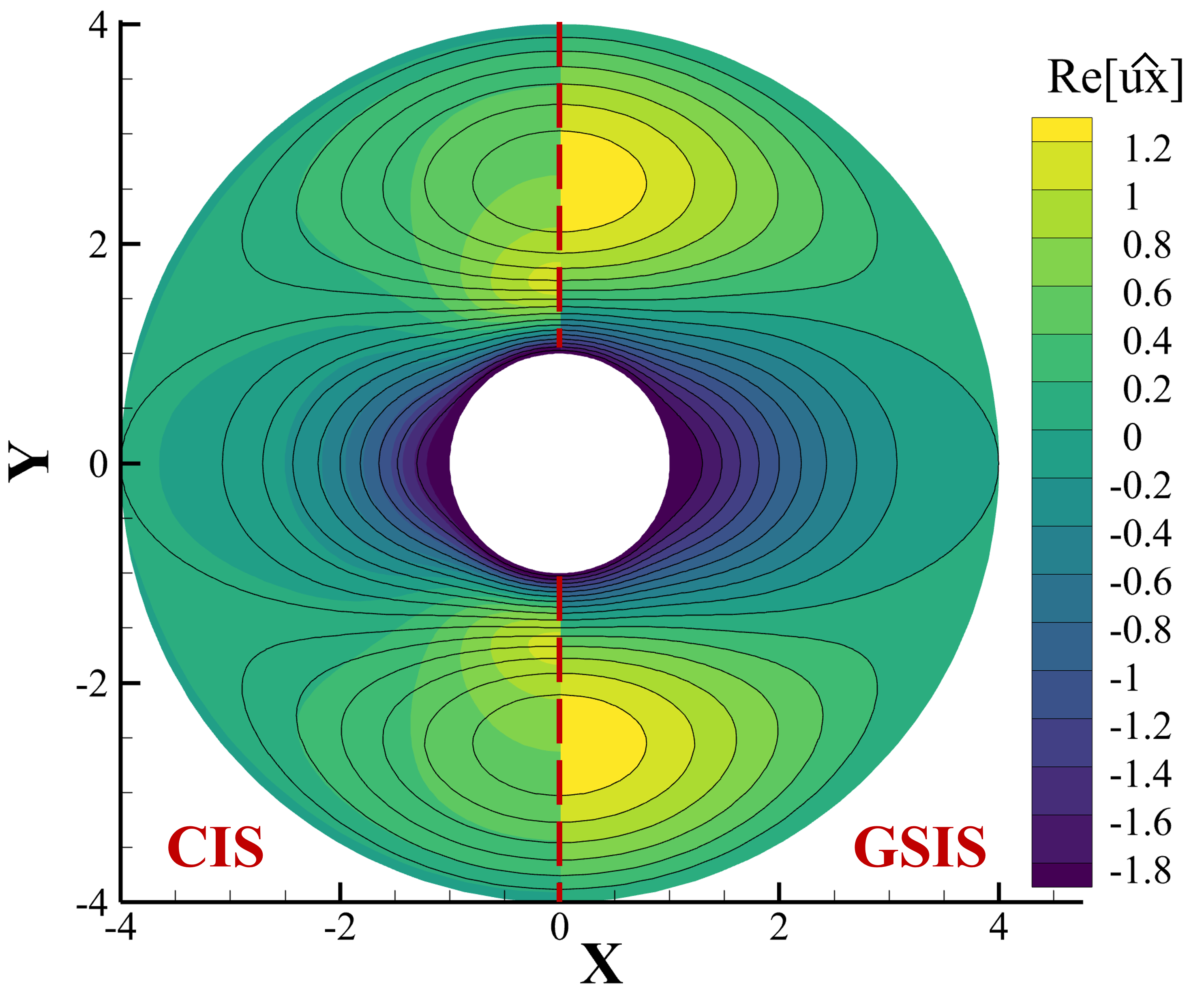}}
    \subfigure[\(\Re(\hat{u}_y)\) when \((\delta_{rp},S)=(1000,10^{-5})\) ]{\includegraphics[height=0.35\linewidth]{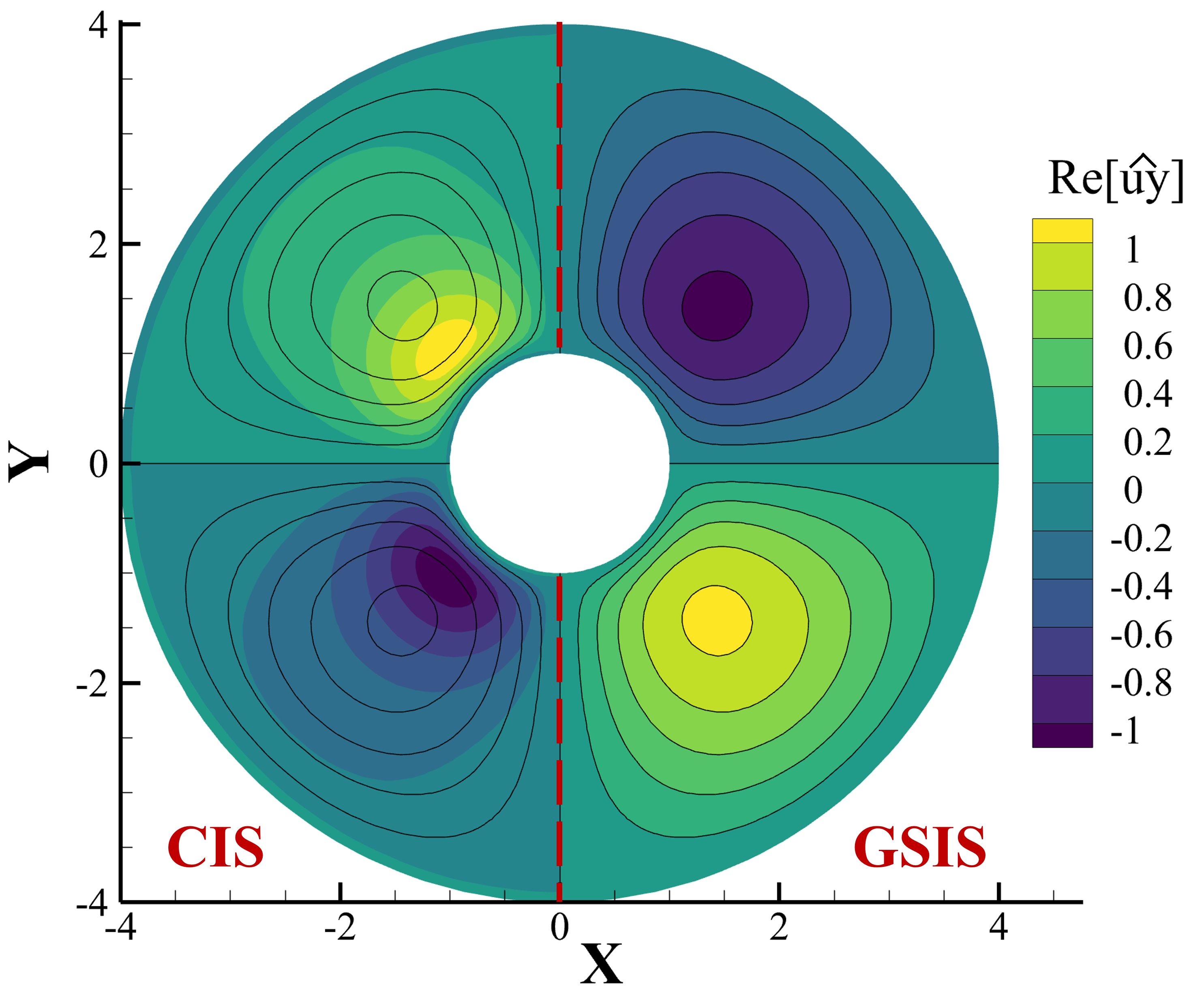}}
    \caption{Geometry, mesh, and adjoint velocity fields for the oscillating-cylinder problem. In the second and third rows, the left and right half contours correspond to CIS and GSIS results, respectively, while the black solid lines represent the reference solutions. 
    }
    \label{fig:cylinder}
\end{figure}

\section{Optimization of an oscillating cylinder}
\label{sec:cylinder_optimization}

The proposed method is first assessed through the shape optimization of an initially circular cylinder. As shown in Fig.~\ref{fig:cylinder}(a), a rarefied gas is confined within the annular region between two concentric cylinders. The radius of the inner cylinder is taken as the reference length $L_{\mathrm{ref}}$, while the radius of the outer cylinder is $4L_{\mathrm{ref}}$. The outer cylinder remains stationary, whereas the inner cylinder undergoes a small-amplitude harmonic oscillation in the horizontal direction. 
In GSIS, the physical domain is discretized using approximately \(2.0\times10^4\) cells, with a minimum near-wall spacing of \(10^{-3}\). A non-uniform grid is adopted for each component $v_\alpha$ ($\alpha\in\{x,y\}$):
\begin{equation}\label{eq:velocity_grid}
    v_{\alpha}^{(k)} = V_{\max}\left(\frac{2k-(N_{v,\alpha}+1)}{N_{v,\alpha}-1}\right)^{3},\quad
    k=1,2,\cdots,N_{v,\alpha},
\end{equation}
which is sufficient to capture the oscillatory structure of the velocity distribution function induced by the vibrating boundary \cite{wu2014JFM,wu2016PRE}. Here we use $N_{v,x}=N_{v,y}=32$ and $V_{\max}=6$.

For both the primal and adjoint solvers, convergence of CIS and GSIS is declared when the following relative residual falls below a prescribed tolerance:
\begin{equation}
    \epsilon_{\mathrm{res}}=\left[\frac{\displaystyle\int_{\Omega}\left\|\bm U^{n+1}-\bm U^n\right\|_2^2\,\mathrm{d}x\,\mathrm{d}y}{\displaystyle\int_{\Omega}\left\|\bm U^n\right\|_2^2\,\mathrm{d}x\,\mathrm{d}y}\right]^{1/2},
\end{equation}
where \(\bm U=[\rho,u_x,u_y,\tau]^\top\) for the primal solver, with an analogous definition for the adjoint solver. Unless otherwise specified, the convergence tolerance is set to \(10^{-6}\).

\subsection{Asymptotic-preserving  and fast convergence}
\label{subsubsec:cylinder_ap}

We assess the asymptotic‑preserving and fast‑convergence properties of the adjoint GSIS.
Figure~\ref{fig:cylinder}(b,c) compares the adjoint velocity fields predicted by CIS and GSIS on identical meshes when \((\delta_{rp},S)=(100,1)\). Taking advantage of the geometric symmetry about \(x=0\), CIS results are plotted in the left computational domain and GSIS results in the right domain. The black solid lines represent the reference solution acquired via CIS on a refined mesh containing approximately \(4.0\times10^4\) cells. Both CIS and GSIS achieve excellent agreement with the reference data, validating the accuracy of the proposed adjoint GSIS.

For the near-continuum flow regime with \((\delta_{rp},S)=(1000,10^{-5})\), the adjoint velocity fields from CIS and GSIS are compared in Figure~\ref{fig:cylinder}(d,e), with the NS solution adopted as the benchmark. The maximum cell size is \(0.13\), which is substantially larger than the mean free path of order \(10^{-3}\), leaving the kinetic scale unresolved over most of the computational domain. Nevertheless, GSIS accurately reproduces the NS benchmark solution, whereas the CIS produces obvious discrepancies. This case demonstrates the prominent asymptotic-preserving property of the developed adjoint GSIS.

\begin{table}[h]
\centering
\caption{Convergence performance of CIS and GSIS for solving the
adjoint equations of the oscillating-cylinder problem. The code is implemented in double precision with OpenMP parallelization and executed on AMD EPYC 7763 processor (2.45 GHz) using 8 threads.
}
\label{tab:cylinder-cost}
\renewcommand{\arraystretch}{1.05}
\begin{tabular}{ccccccc}
\toprule
\multirow{2}{*}{$(\delta_{rp},S)$} &
\multirow{2}{*}{$N_{\text{cell}}$} &
\multirow{2}{*}{$N_{\text{vel}}$} &
\multicolumn{2}{c}{Iteration steps} &
\multicolumn{2}{c}{Wall-clock time (s)} \\
\cmidrule(lr){4-5}\cmidrule(lr){6-7}
 &   &  & CIS & GSIS & CIS & GSIS \\
\midrule
\((100,1.0)\)     & 20,000 & 1024  & 27\,482 & 28 & 23\,984  & 68 \\
\((1000,10^{-5})\) & 20,000 & 1024 & 100,000 & 36 & 69,595 & 88 \\
\bottomrule
\end{tabular}
\end{table}

Convergence statistics listed in Table~\ref{tab:cylinder-cost} further reveal that GSIS converges within merely a few tens of iterations and cuts the wall‑clock time by several orders of magnitude for the near‑continuum flow. This demonstrates that the fast-convergence capability of the adjoint GSIS, originally validated via Fourier stability analysis on infinite domains in Section \ref{subsec:adjoint_gsis_convergence}, persists even for wall-bounded flows.

\subsection{Optimization setup and sensitivity verification}

The moving boundary is parameterized using the FFD formulation introduced in Section~\ref{sec:ffd}. The initial cylinder is embedded in a square FFD lattice spanning \([-1.5,1.5]^2\), with \(8\times8\) uniformly distributed control points and cubic B-spline basis functions in both parametric directions.

The objective is to reduce the amplitude of the complex horizontal gas-force response by modifying the shape of the inner cylinder. Let \(J(\bm Q)\) denote the complex amplitude of the horizontal force exerted on the moving wall. 
The optimization problem is formulated as
\begin{equation}
\begin{aligned}
    \min_{\bm Q}\quad&
    J_{\mathrm{obj}}(\bm Q)
    =
    \left|J(\bm Q)\right|^2,
    \quad
    \text{subject to} \quad
    A(\bm Q^0)-A(\bm Q)\leq 0.
\end{aligned}
\label{eq:ba_optimization}
\end{equation}
Here, \(A\) is the area of the inner cylinder and $\bm Q^0$ is the parameter for the initial circular cylinder. It's worth to note that,  for this inequality constraint, we should calculate \(\partial A/\partial\bm Q\) to pass to the optimizer.

\begin{figure}[t]
    \centering
    \includegraphics[height=0.35\linewidth]{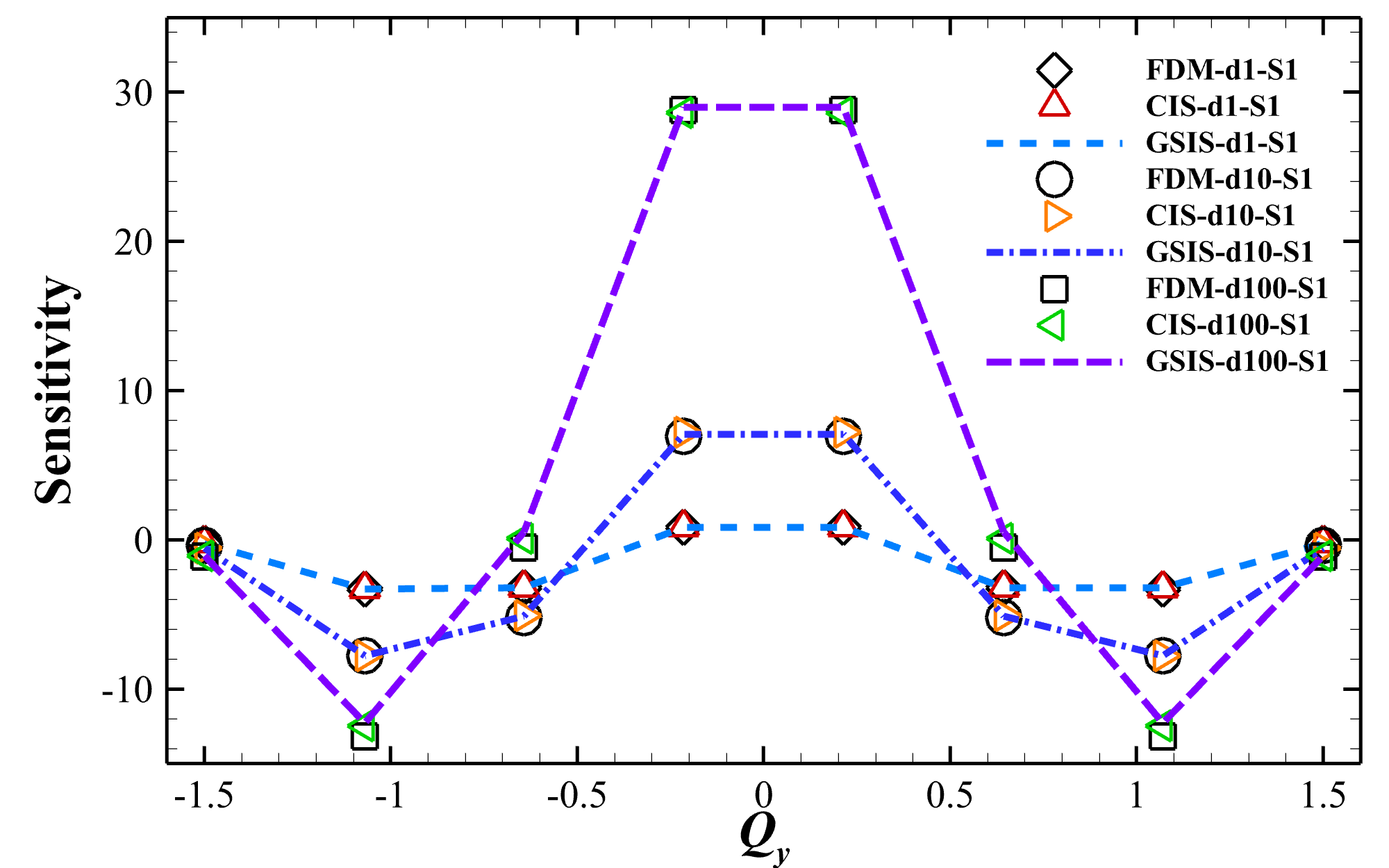}
    \caption{The sensitivity \(\partial J_{\text{obj}}/\partial Q_x\) for the oscillating cylinder. In the legend, the numerical suffixes following "d" and "S" denote the rarefaction parameter \(\delta_{rp}\) and Strouhal number \(S\), respectively.
    }
    \label{fig:cylinder_sensitivity}
\end{figure}

To verify the gradient supplied to the optimizer, the objective sensitivity is compared with that obtained by the finite-difference method (FDM). Let $(\bm{Q}_{ij})_q$ denote the $q$-direction coordinate of the FFD control point $\bm{Q}_{ij}$, where $q\in\{x,y\}$. Its finite-difference sensitivity is evaluated using the central-difference formula
\begin{equation}
    \left(
    \frac{J_{\text{obj}}}{\partial (\bm{Q}_{ij})_q}
    \right)_{\mathrm{FDM}}
    =
    \frac{
        J_{\text{obj}}\!\left[(\bm{Q}_{ij})_q+\epsilon\right]
        - J_{\text{obj}}\!\left[(\bm{Q}_{ij})_q-\epsilon\right]
    }{2\epsilon},
    \label{eq:fdm_sensitivity}
\end{equation}
where the perturbation amplitude is set to $\epsilon=10^{-3}$.

Figure~\ref{fig:cylinder_sensitivity} compares \(\partial J_{\text{obj}}/\partial Q_x\) obtained by FDM, CIS, and GSIS for $(\delta_{rp},S)=(1,1)$, $(10,1)$, and $(100,1)$. The comparison is performed for the FFD control points whose  horizontal coordinate is $Q_x=-0.64$. The GSIS sensitivities agree closely with the CIS and FDM results, confirming the accuracy of the adjoint GSIS.

\begin{figure}[t]
    \centering
    \includegraphics[width=0.5\linewidth]{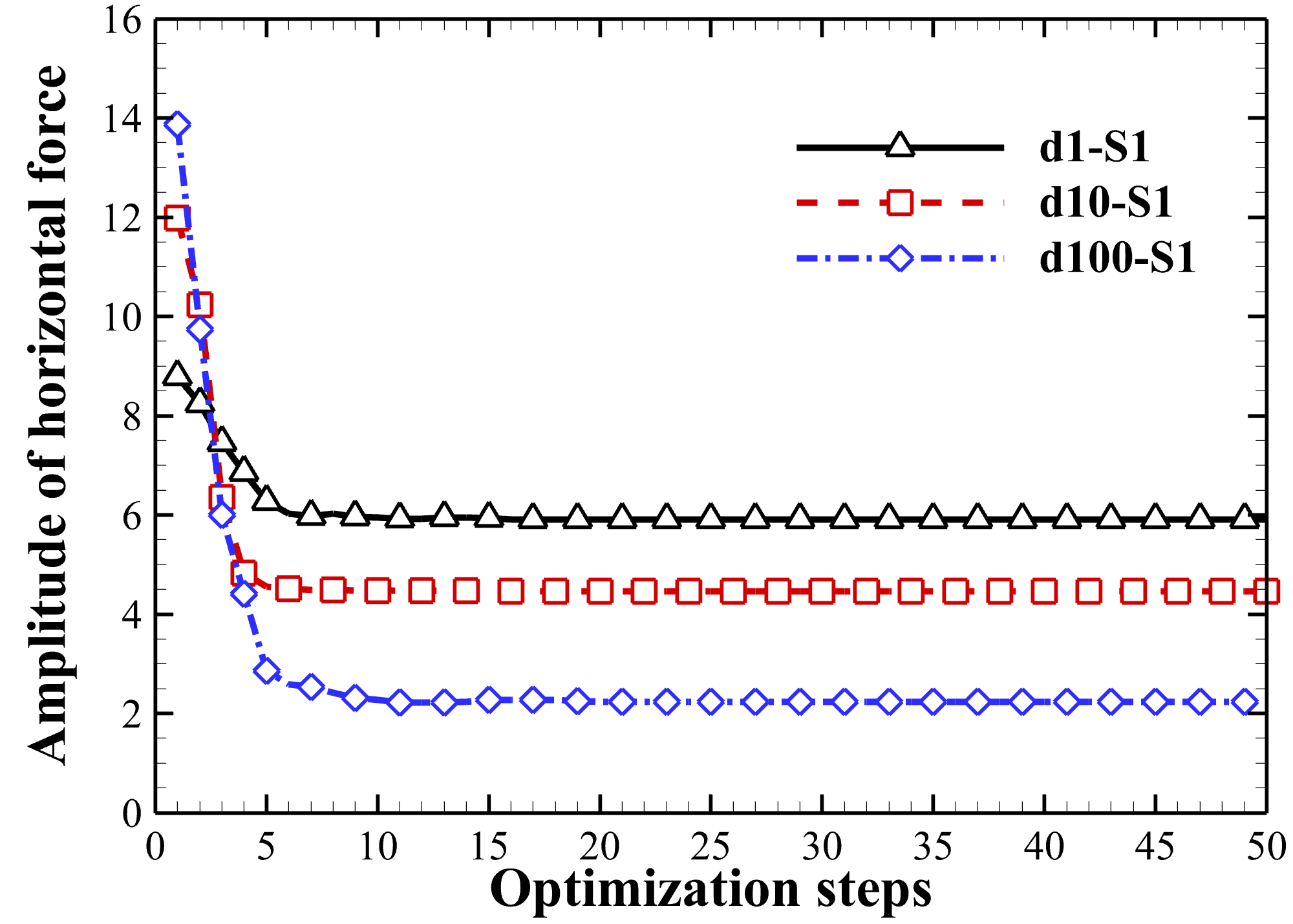}
    \caption{Optimization history for the oscillating cylinder.
    }
    \label{fig:cylinder_obj_contribution}
\end{figure}

\begin{table}[t]
    \centering
    \caption{Complex horizontal force \(J\) at different stages of the oscillating-cylinder optimization. The reduction is calculated from \(|J|\) relative to the initial circular geometry.}
    \label{tab:cylinder_force_reduction}
    \small
    \renewcommand{\arraystretch}{1.15}
    \setlength{\tabcolsep}{6pt}
    \begin{tabular}{@{}ccccc@{}}
        \toprule
        \((\delta_{rp},S)\)
        & Initial
        & Step 10
        & Final optimized
        & Reduction\\
        \midrule
        \((1,\,1)\)
        & \(-8.66-1.54\mathrm{i}\)
        & \(-5.85-1.07\mathrm{i}\)
        & \(-5.79-1.06\mathrm{i}\)
        & \(33.1\%\) \\
        \((10,\,1)\)
        & \(-10.9-4.99\mathrm{i}\)
        & \(-3.82-2.34\mathrm{i}\)
        & \(-3.80-2.33\mathrm{i}\)
        & \(62.8\%\) \\
        \((100,\,1)\)
        & \(-5.65-12.7\mathrm{i}\)
        & \(-1.33-1.85\mathrm{i}\)
        & \(-1.33-1.79\mathrm{i}\)
        & \(84.0\%\) \\
        \bottomrule
    \end{tabular}
\end{table}

\subsection{Optimization results}

Finally, shape optimizations are performed for \((\delta_{rp},S)=(1,1)\), \((10,1)\), and \((100,1)\). Figure~\ref{fig:cylinder_obj_contribution} shows that the horizontal force amplitude decreases rapidly during the first several optimization steps and then approaches a plateau. As listed in Table~\ref{tab:cylinder_force_reduction}, the reductions after ten steps are already \(32.4\%\), \(62.6\%\), and \(83.6\%\), close to the final values of \(33.1\%\), \(62.8\%\), and \(84.0\%\), respectively. Thus, most of the force reduction is achieved within approximately ten optimization steps. The optimization reduces both the real and imaginary components of \(J\), with the overall reduction increasing markedly as \(\delta_{rp}\) increases. In particular, the initially dominant imaginary component at \(\delta_{rp}=100\) is strongly suppressed.

\begin{figure}[t!]
    \centering
    \includegraphics[width=0.4\linewidth]{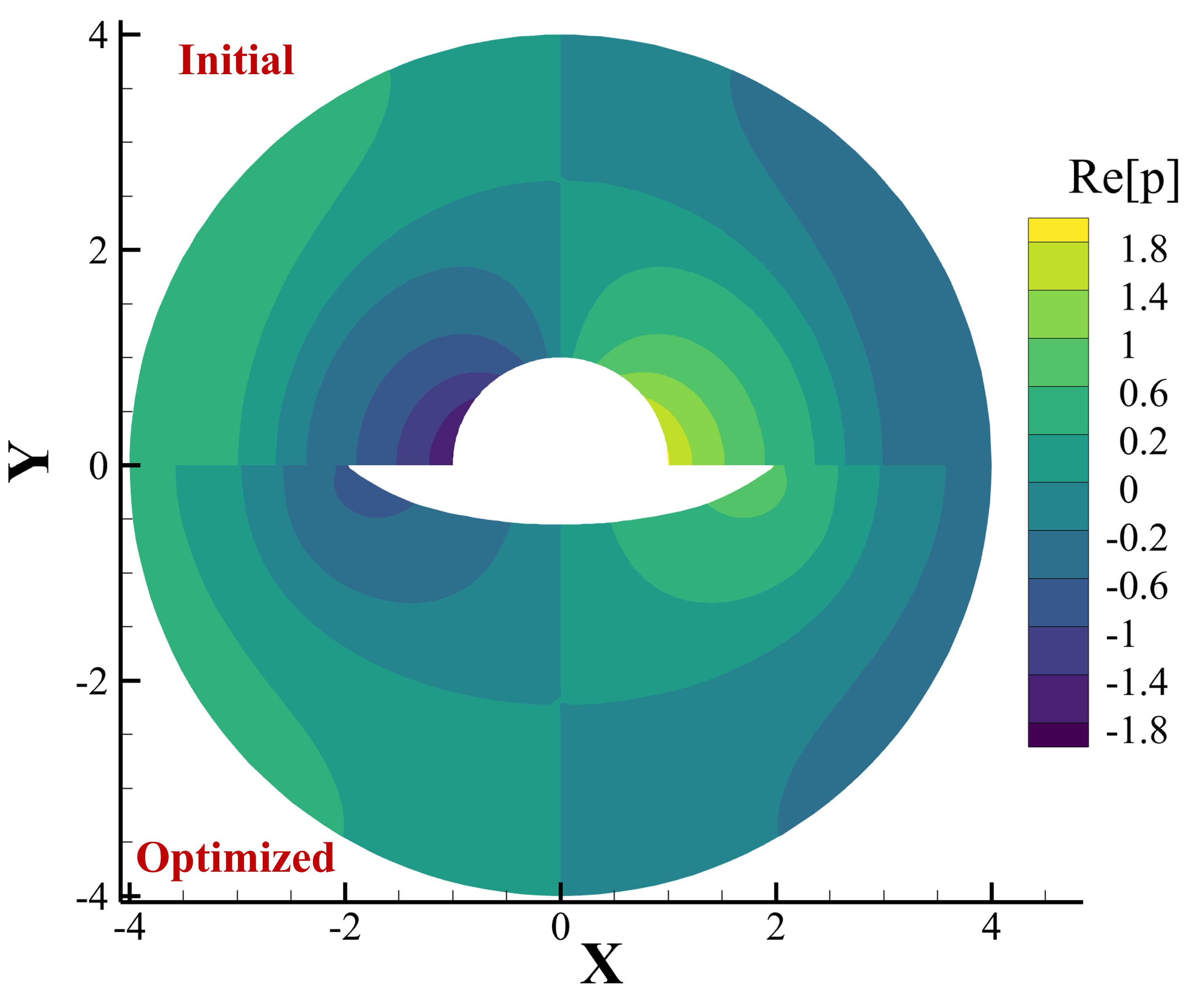}
    \includegraphics[width=0.4\linewidth]{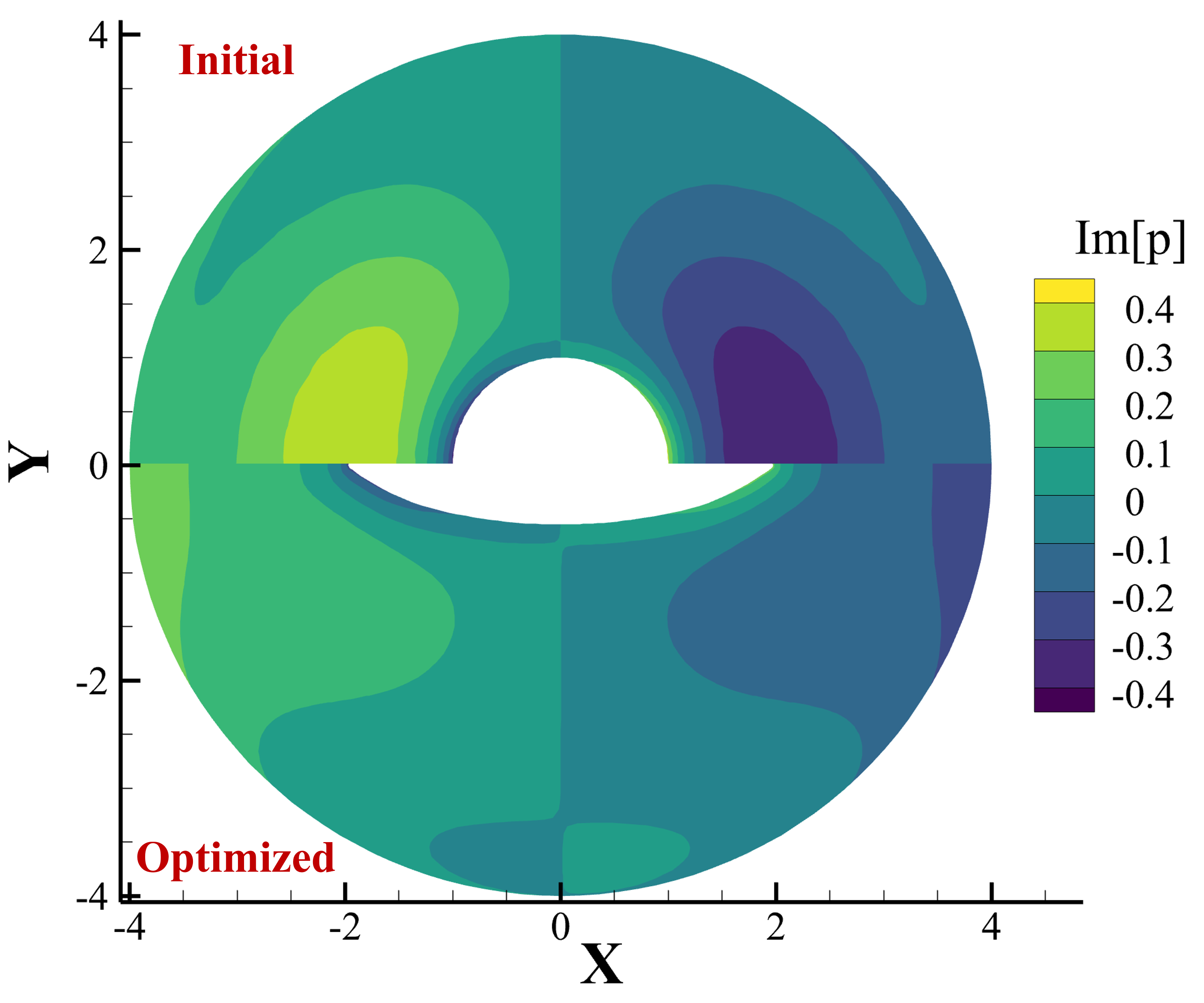}   
    \includegraphics[width=0.4\linewidth]{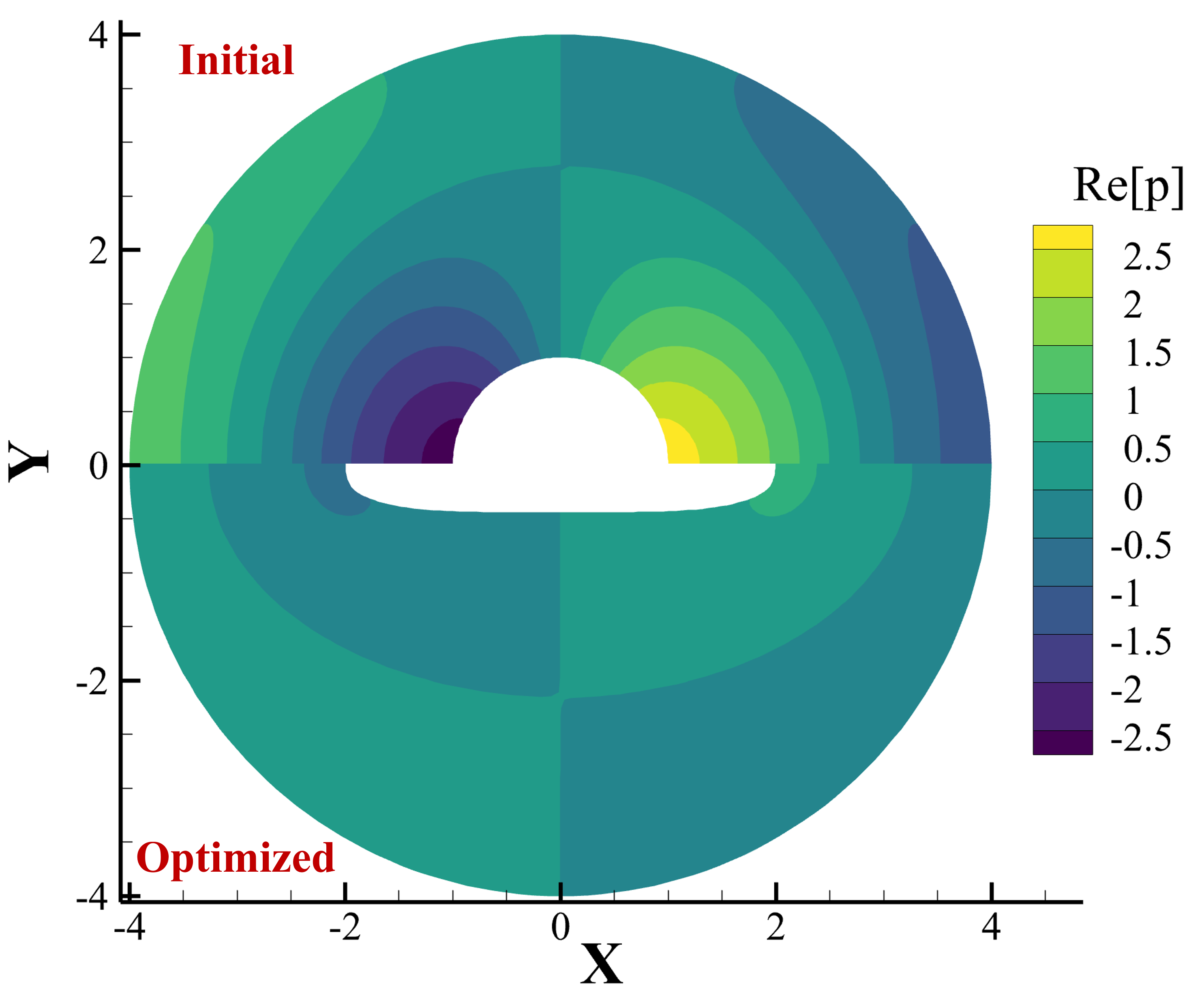}
    \includegraphics[width=0.4\linewidth]{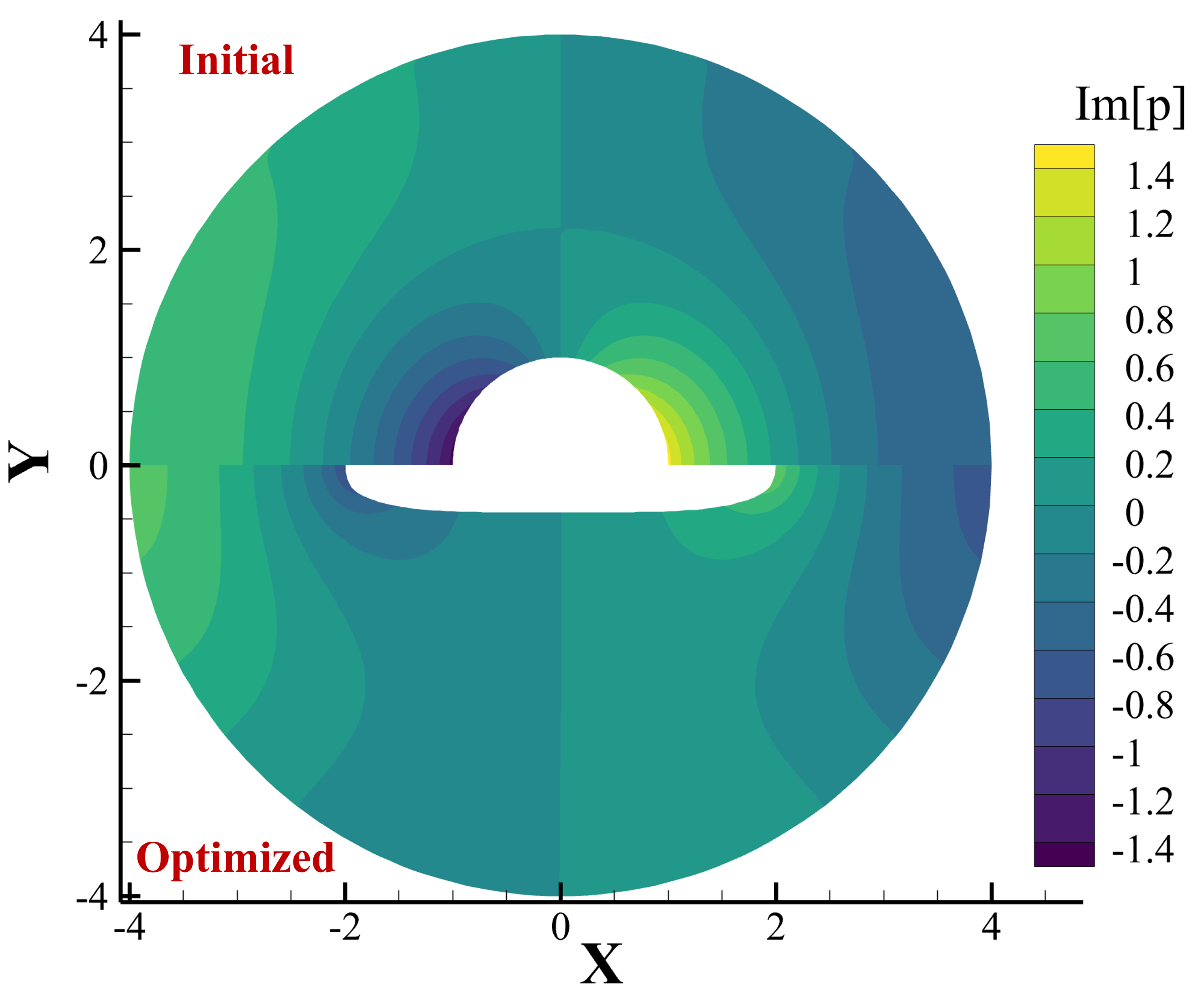}  
    \includegraphics[width=0.4\linewidth]{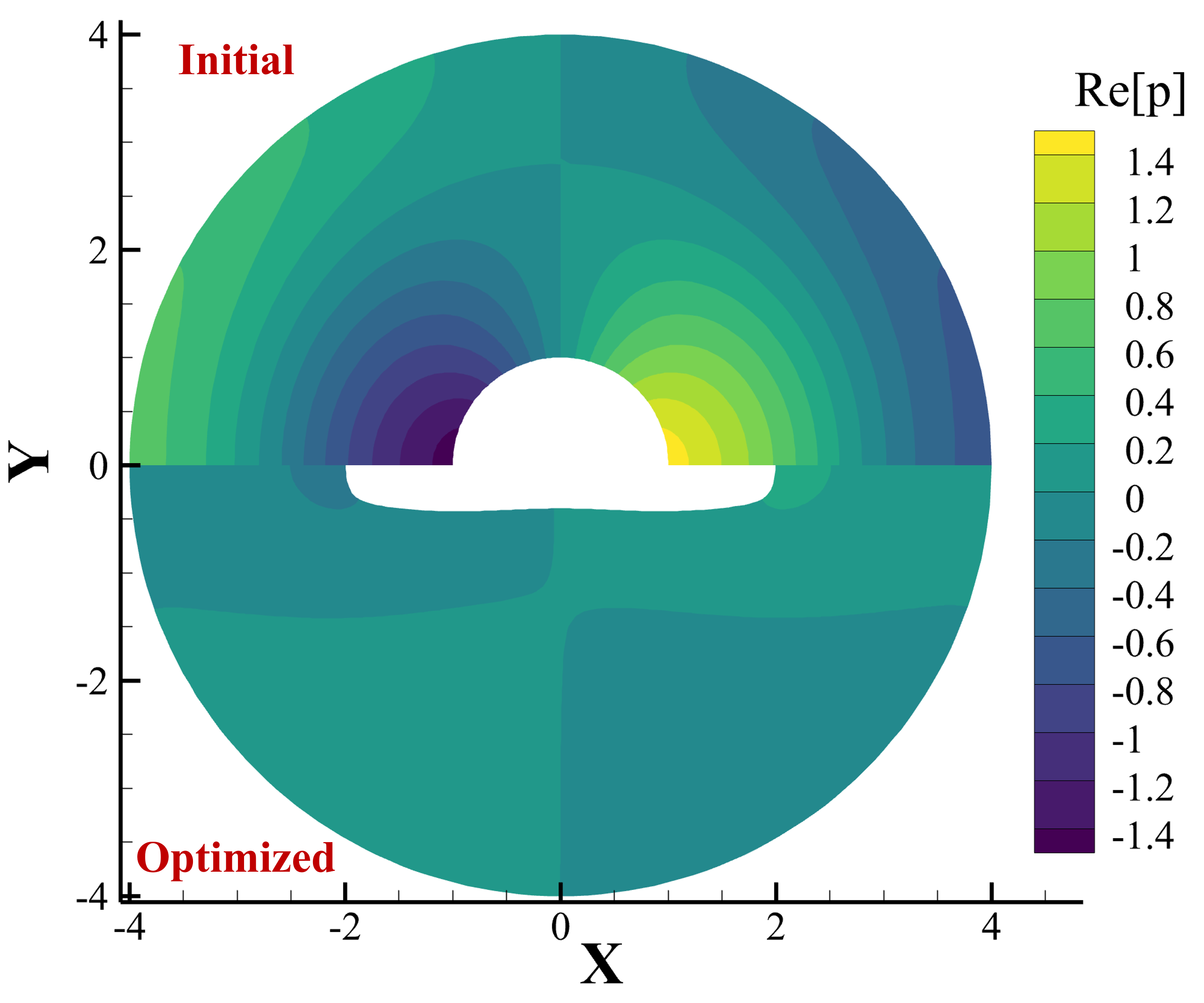}
    \includegraphics[width=0.4\linewidth]{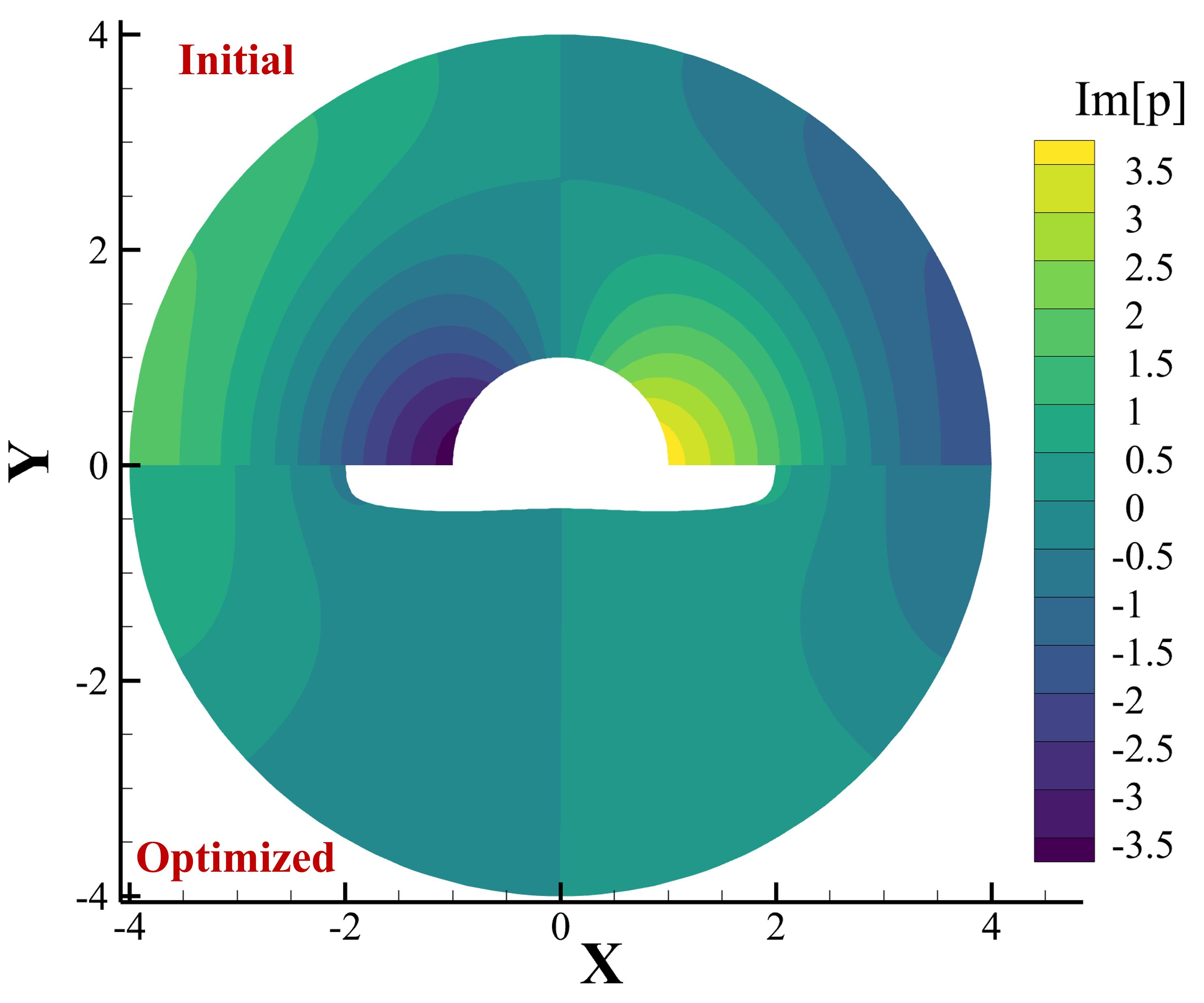}  
    \caption{Initial and optimized oscillating‑cylinder shapes. Rarefaction parameters are \(\delta_{rp}=1\), 10, and 100 from top to bottom, with Strouhal number \(S=1\). Owing to symmetry, each plot displays pressure contours for the initial geometry in its upper half and for the optimized geometry in its lower half.
    }
    \label{fig:cylinder_shape_results}
\end{figure}

The optimized shapes and the corresponding pressure perturbations are shown in Fig.~\ref{fig:cylinder_shape_results}. Owing to symmetry about the \(x\)-axis, only one half of the solution is displayed for each geometry, with the upper and lower half-domains corresponding to the initial and optimized configurations, respectively. The pressure perturbation exhibits opposite signs on the two sides of the cylinder, reflecting the pressure imbalance responsible for the horizontal force. After optimization, the high-amplitude pressure regions are substantially weakened and confined to the leading and trailing portions of the elongated body, thereby reducing the integrated force response. This suppression is particularly pronounced for the imaginary component at \(\delta_{rp}=100\), consistent with the complex-force results in Table~\ref{tab:cylinder_force_reduction}.

For all three rarefaction parameters, the optimized cylinder is flattened in the transverse direction and elongated along the oscillation direction. At \(\delta_{rp}=1\), the optimized profile retains a relatively thick central region and develops sharper leading and trailing edges. By contrast, the optimized geometries for \(\delta_{rp}=10\) and \(100\) are more elongated and exhibit similar smooth profiles, suggesting a gradual transition toward a common optimal configuration.

\section{Optimization of a biaxial accelerometer}
\label{sec:ba}

\begin{figure}[t!]
    \centering
    \subfigure[]{\includegraphics[height=0.29\linewidth]{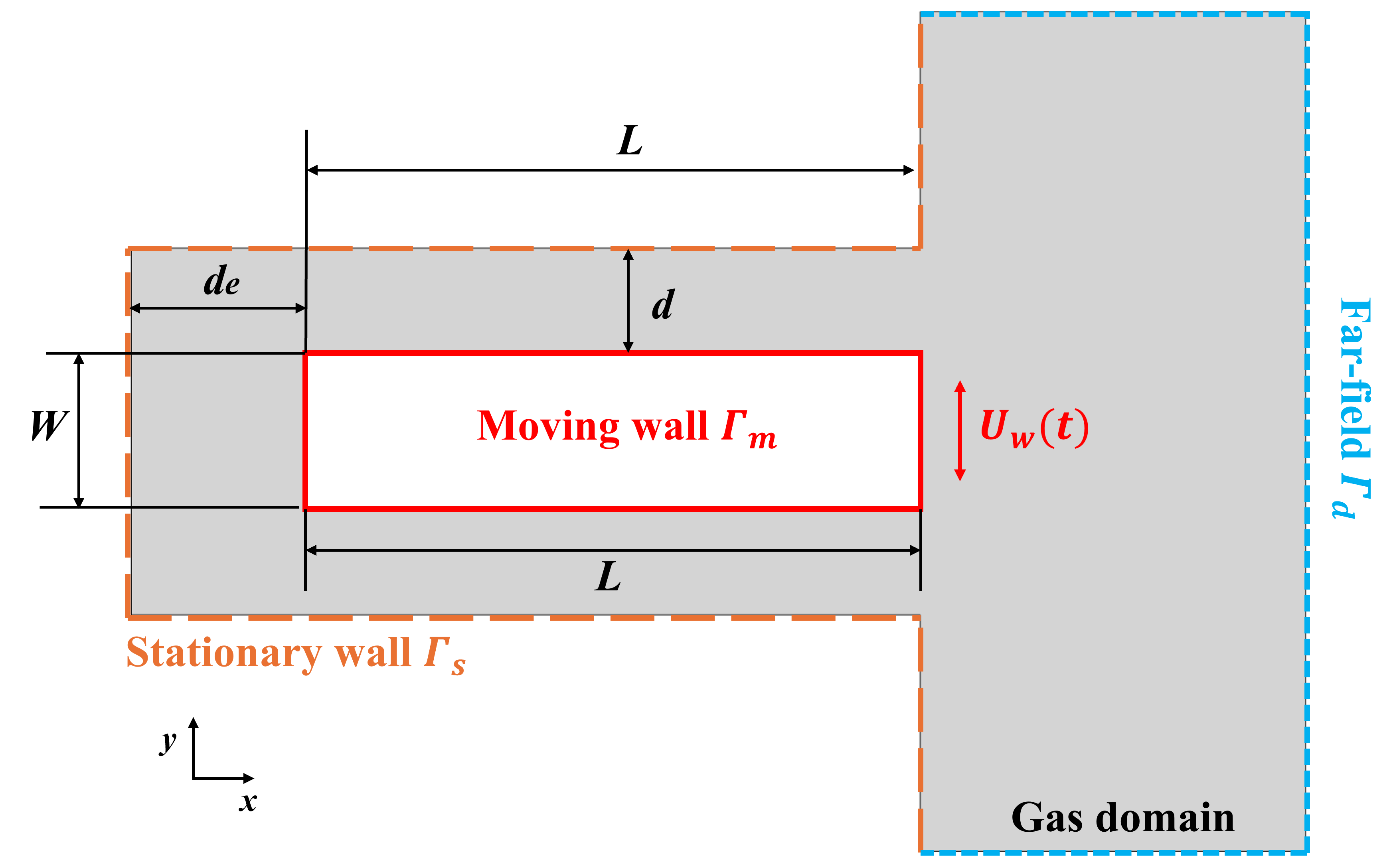}}
    \hspace{0.3cm}
    \subfigure[]{\includegraphics[height=0.29\linewidth]{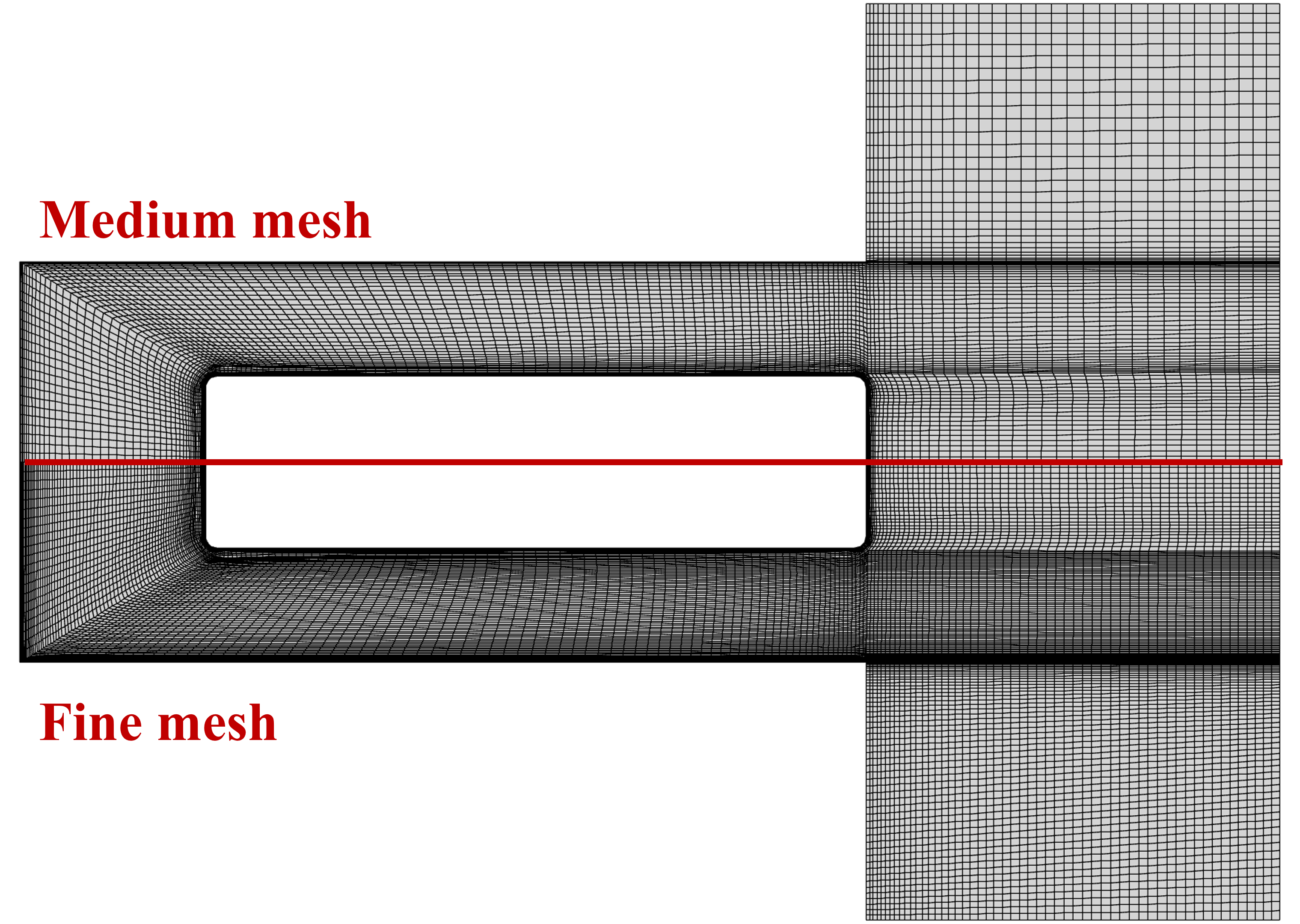}} \\
    \subfigure[]{\includegraphics[width=0.4\linewidth]{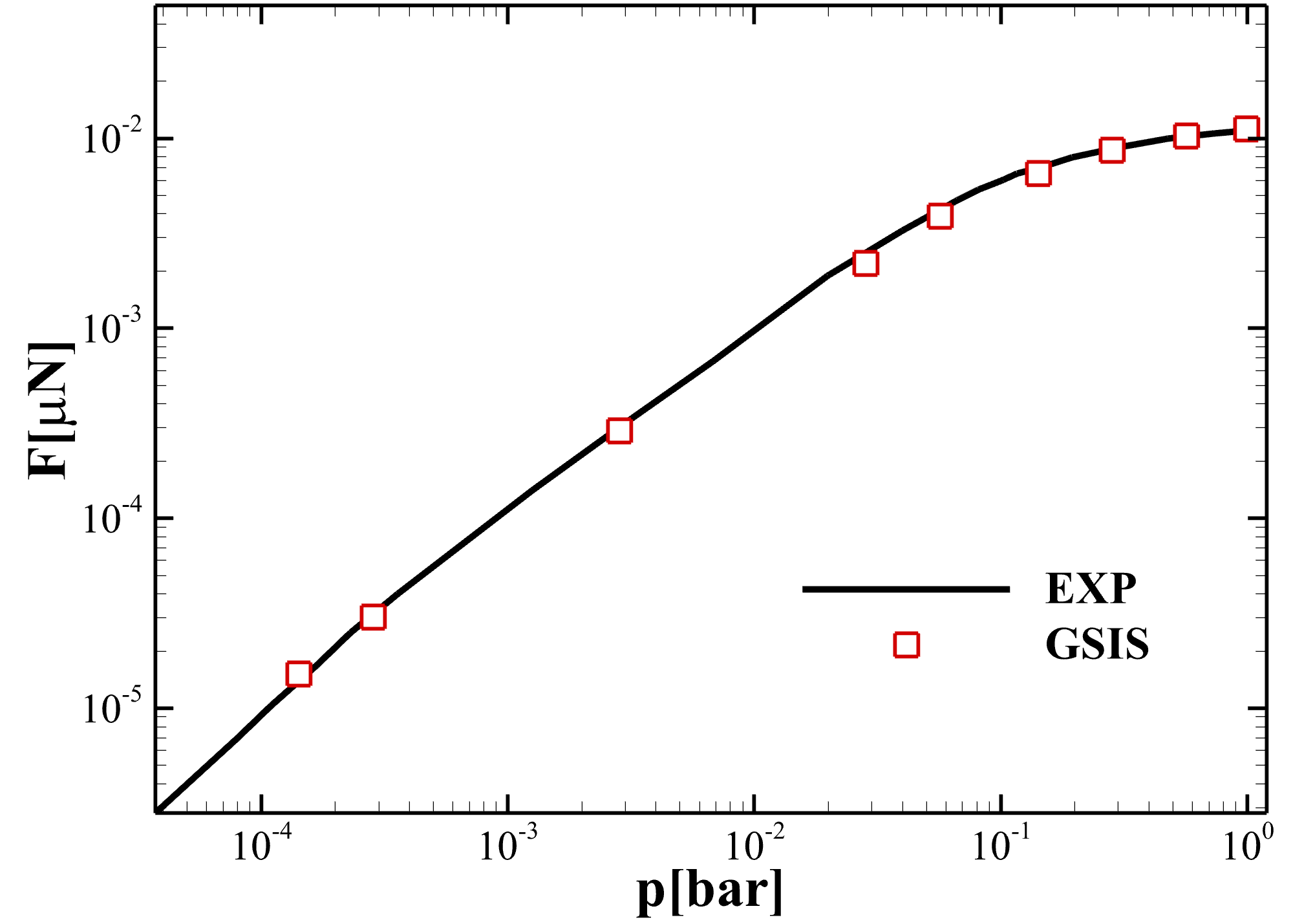}}
    \caption{(a) Two-dimensional schematic of a biaxial accelerometer. 
    (b) Computational mesh. The upper half displays the medium‑resolution mesh with \(18\,148\) cells and a minimum near‑wall spacing of \(10^{-3}L_{\mathrm{ref}}\), whereas the lower half presents the fine mesh of approximately \(3.7\times10^4\) cells, obtained by further refining the narrow gaps and outer domain.
    (c) Damping force from the GSIS and experiment~\cite{frangi2007application}. 
    }
    \label{fig:ba_mesh}
\end{figure}

The proposed method is further applied to the biaxial accelerometer~\cite{frangi2007application}. Owing to the large aspect ratio of the interdigitated plates, the gas flow around a representative shuttle--stator pair is modeled using the two-dimensional cross-section shown in Fig.~\ref{fig:ba_mesh}(a). The movable shuttle boundary \(\Gamma_m\) undergoes a small-amplitude harmonic translation in the \(y\) direction, whereas the stator boundary \(\Gamma_s\) remains stationary. Diffuse reflection is imposed on all solid surfaces, and the equilibrium distribution is prescribed for the far-field boundary \(\Gamma_d\). Originally, the shuttle height and thickness are \(L=15~\mu\mathrm{m}\) and \(W=3.9~\mu\mathrm{m}\), respectively, while the shuttle--stator gap and the clearance from the substrate are \(d=2.6~\mu\mathrm{m}\) and \(d_e=4.2~\mu\mathrm{m}\), respectively. The reference length is chosen as \(L_{\mathrm{ref}}=2.6~\mu\mathrm{m}\).

We first assess the accuracy of the primal solver. 
In numerical simulations, the \(32\times32\) nonuniform velocity grid is used as per Eq.~\eqref{eq:velocity_grid}. As shown in Fig.~\ref{fig:ba_mesh}(b), the medium spatial mesh is used. Following Ref.~\cite{frangi2007application}, the wall-velocity amplitude is set to \(U_0=1~\mathrm{m/s}\), and an out-of-plane depth of \(b=1~\mu\mathrm{m}\) is assumed. The dimensional force acting on the representative shuttle cross-section is evaluated as $F=-\xi p_0 (bL_{\mathrm{ref}})\,\Re\{J\}$,
where the minus sign indicates that the damping force opposes the wall motion. Good agreement against the experimental data in Fig.~\ref{fig:ba_mesh}(c) validates the accuracy of the present GSIS solver.

\subsection{Optimization setup and sensitivity verification}
\label{sec:ba_opt_setup}

The optimization problem is formulated as
\begin{equation}
\begin{aligned}
    \min_{\bm Q}\quad & J_{\mathrm{obj}}(\bm Q)=\left|J(\bm Q)\right|^2,\\
    \text{subject to}\quad & A(\bm Q^0)-A(\bm Q)\leq 0,\\
    & \frac{\left\|\bm Q_{i-1,j}-2\bm Q_{ij}+\bm Q_{i+1,j}\right\|_{\infty}}{h_u}\leq\eta,\\
    & \frac{\left\|\bm Q_{i,j-1}-2\bm Q_{ij}+\bm Q_{i,j+1}\right\|_{\infty}}{h_v}\leq\eta,\\
    & 0\leq (\bm{Q}_{ij})_x\leq 7.5,\qquad -1.2\leq (\bm{Q}_{ij})_y\leq 1.2,
\end{aligned}
\label{eq:ba_optimization}
\end{equation}
where \(A\) denotes the cross-sectional area of the movable shuttle, \(h_u\) and \(h_v\) are the control-point spacings in the two parametric directions. The area constraint prevents force reduction through structural shrinkage. The parameter \(\eta\), set to \(0.4\), limits the normalized second-order differences of adjacent control points and thereby suppresses excessive local curvature. A smaller value would overly restrict the design space and keep the optimized geometry close to the initial shape. The coordinate bounds restrict the admissible deformation region, preventing excessive mesh distortion and nonphysical geometries.

The design boundary of the movable shuttle is parameterized using the FFD formulation. The shuttle is embedded in a rectangular FFD lattice spanning \(x\in[1.6,7.4]\) and \(y\in[-0.76,0.76]\), with \(6\times6\) uniformly distributed control points and cubic B-spline basis functions in both parametric directions. Since the flow configuration is symmetric about the \(x\)-axis, only the displacements of the 18 control points in the lower half of the lattice are treated as independent design variables, while those in the upper half are determined by mirror symmetry. To avoid non-smooth boundary updates and mesh deterioration near the sharp corners, the original shuttle is rounded with a fillet radius \(r=0.1L_{\mathrm{ref}}\), and the resulting geometry is used as the initial design.

\begin{figure}[t!]
    \centering
    \subfigure[]{\includegraphics[width=0.45\linewidth]{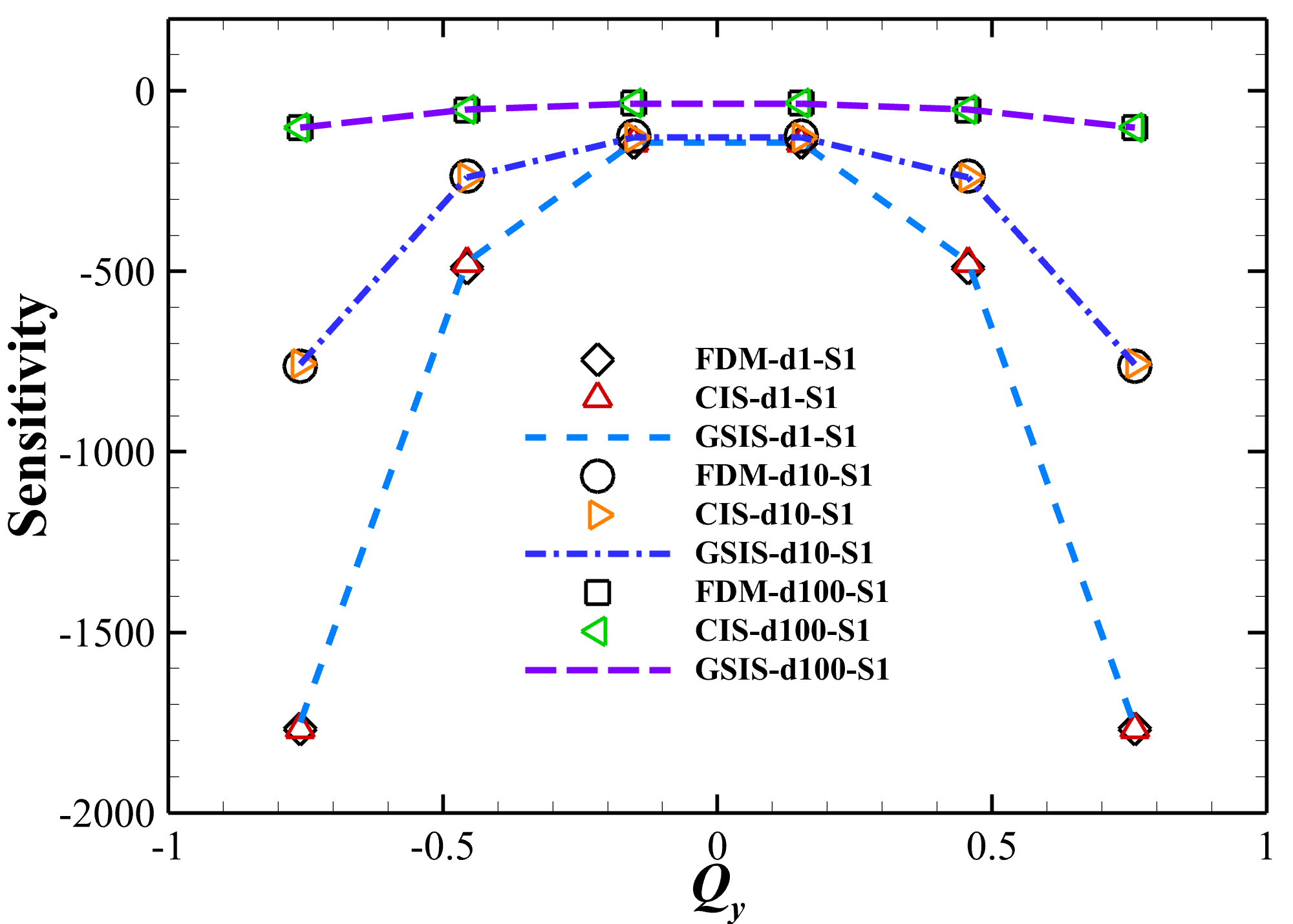}}
        \hspace{0.2cm}
    \subfigure[]{\includegraphics[width=0.45\linewidth]{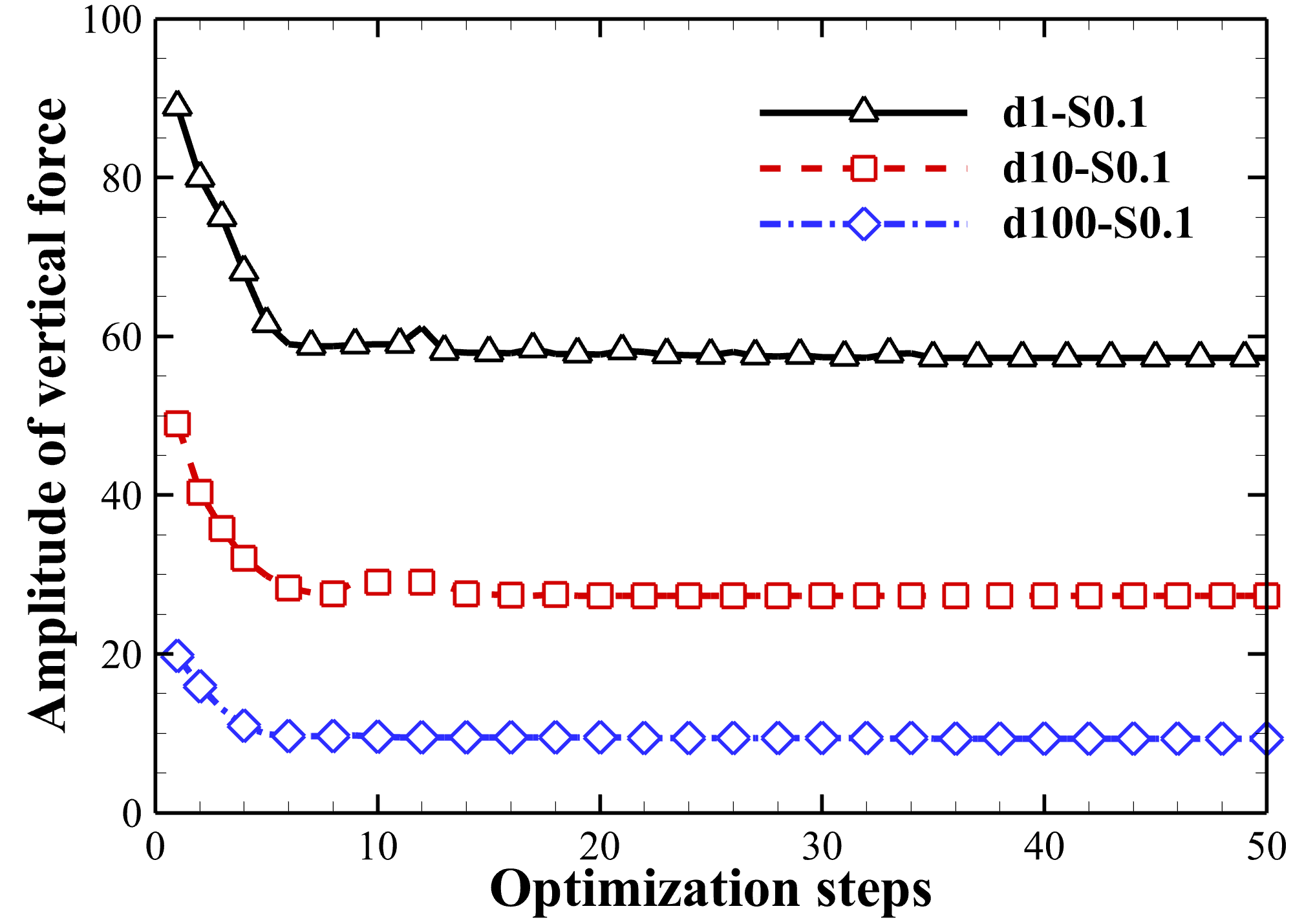}}
    \caption{
    Biaxial accelerometer.
    (a) Comparison of selected FFD sensitivities \(\partial J_{\text{obj}}/\partial Q_x\) obtained by FDM, CIS, and GSIS.
    (b) Optimization histories of \(|J|\) for \((\delta_{rp},S)=(1,0.1)\), \((10,0.1)\) and \((100,0.1)\).
    }
    \label{fig:ba}
\end{figure}

A mesh-convergence check of the shape sensitivity is performed at the control point \((Q_x,Q_y)=(1.6,-0.76)\), which exhibits a relatively large sensitivity magnitude under all operating conditions. The relative differences between the medium- and fine-mesh results are below \(1.0\%\) in all cases, indicating that the sensitivity is sufficiently resolved on the medium mesh. The medium mesh is therefore adopted for the subsequent calculations. The adjoint sensitivities obtained using CIS and GSIS are then validated against finite-difference results. Figure~\ref{fig:ba}(a) presents \(\partial J_{\text{obj}}/\partial Q_x\) for the control points in the FFD column \(Q_x=1.6\) as a function of \(Q_y\). The close agreement among the three results confirms the accuracy of the adjoint sensitivity evaluation.


\subsection{Optimization results and computational efficiency}
\label{subsec:ba_results}

Finally, shape optimizations are performed for \((\delta_{rp},S)=(1,0.1)\), \((10,0.1)\), and \((100,0.1)\). As shown in Fig.~\ref{fig:ba}(b), the force amplitude decreases rapidly and then approaches a plateau. Relative to the rounded initial geometry, the final reductions are \(35.5\%\), \(44.3\%\), and \(52.6\%\), respectively, of which more than \(87\%\) is achieved within the first ten steps. Table~\ref{tab:ba_force_reduction} further shows that corner rounding changes the force magnitude by only \(4.8\%\), \(4.5\%\), and \(2.7\%\), confirming that the reductions mainly result from shape optimization rather than geometric pre-processing.

\begin{table}[p]
    \centering
    \caption{Complex force \(J\) at different stages of the biaxial-accelerometer optimization. The
    drag reduction is calculated from \(|J|\) relative to
    the rounded initial geometry.
    }
    \label{tab:ba_force_reduction}
    \small
    \renewcommand{\arraystretch}{1.15}
    \setlength{\tabcolsep}{6pt}
    \begin{tabular}{@{}cccccc@{}}
        \toprule
        \((\delta_{rp},S)\)
        & Original
        & Rounded initial
        & Step 10
        & Final optimized
        & Reduction\\
        \midrule
        \((1,\,0.1)\)
        & \(-71.0+60.6\mathrm{i}\)
        & \(-72.4+51.5\mathrm{i}\)
        & \(-57.4+19.4\mathrm{i}\)
        & \(-54.9+16.4\mathrm{i}\)
        & \(35.5\%\) \\
        \((10,\,0.1)\)
        & \(-51.1+4.65\mathrm{i}\)
        & \(-48.9+3.16\mathrm{i}\)
        & \(-29.8-3.46\mathrm{i}\)
        & \(-27.1-3.16\mathrm{i}\)
        & \(44.3\%\) \\
        \((100,\,0.1)\)
        & \(-10.4-17.4\mathrm{i}\)
        & \(-10.0-17.0\mathrm{i}\)
        & \(-5.63-7.95\mathrm{i}\)
        & \(-5.54-7.53\mathrm{i}\)
        & \(52.6\%\) \\
        \bottomrule
    \end{tabular}
\end{table}

\begin{figure}[p]
    \centering
    \includegraphics[width=0.45\linewidth]{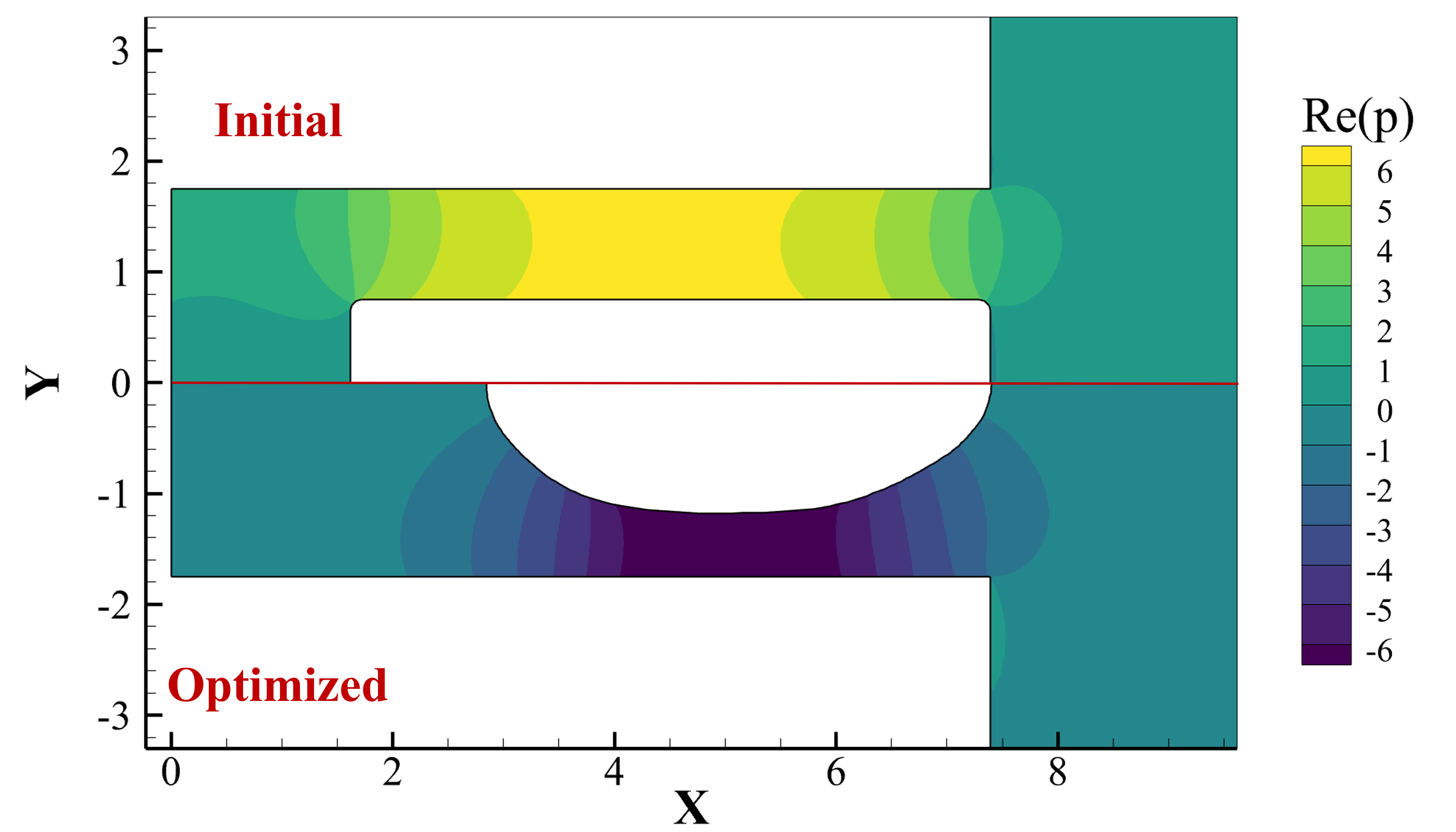}
    \hspace{0.1cm}
    \includegraphics[width=0.45\linewidth]{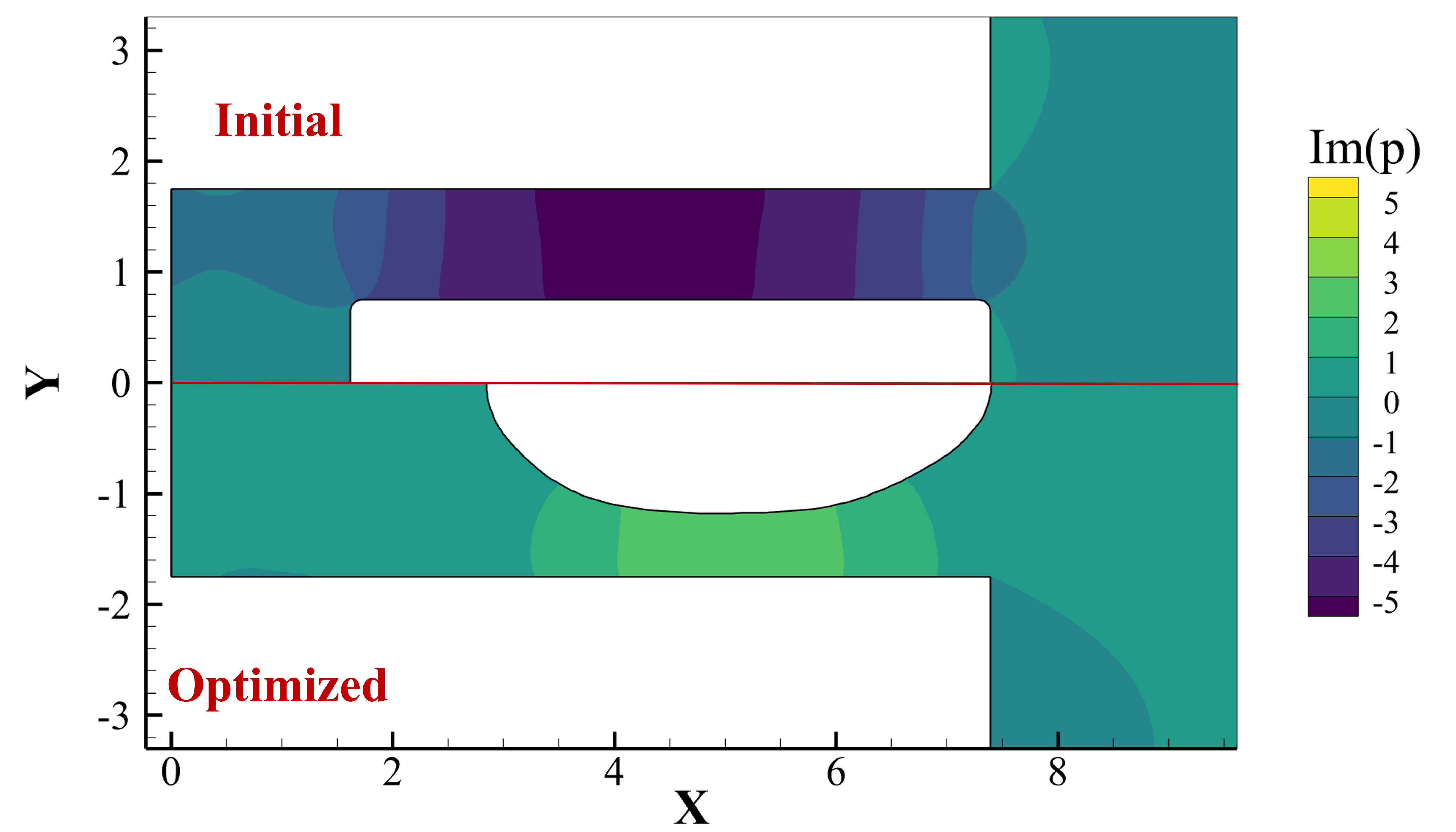}\\
    \vspace{0.3cm}
    \includegraphics[width=0.45\linewidth]{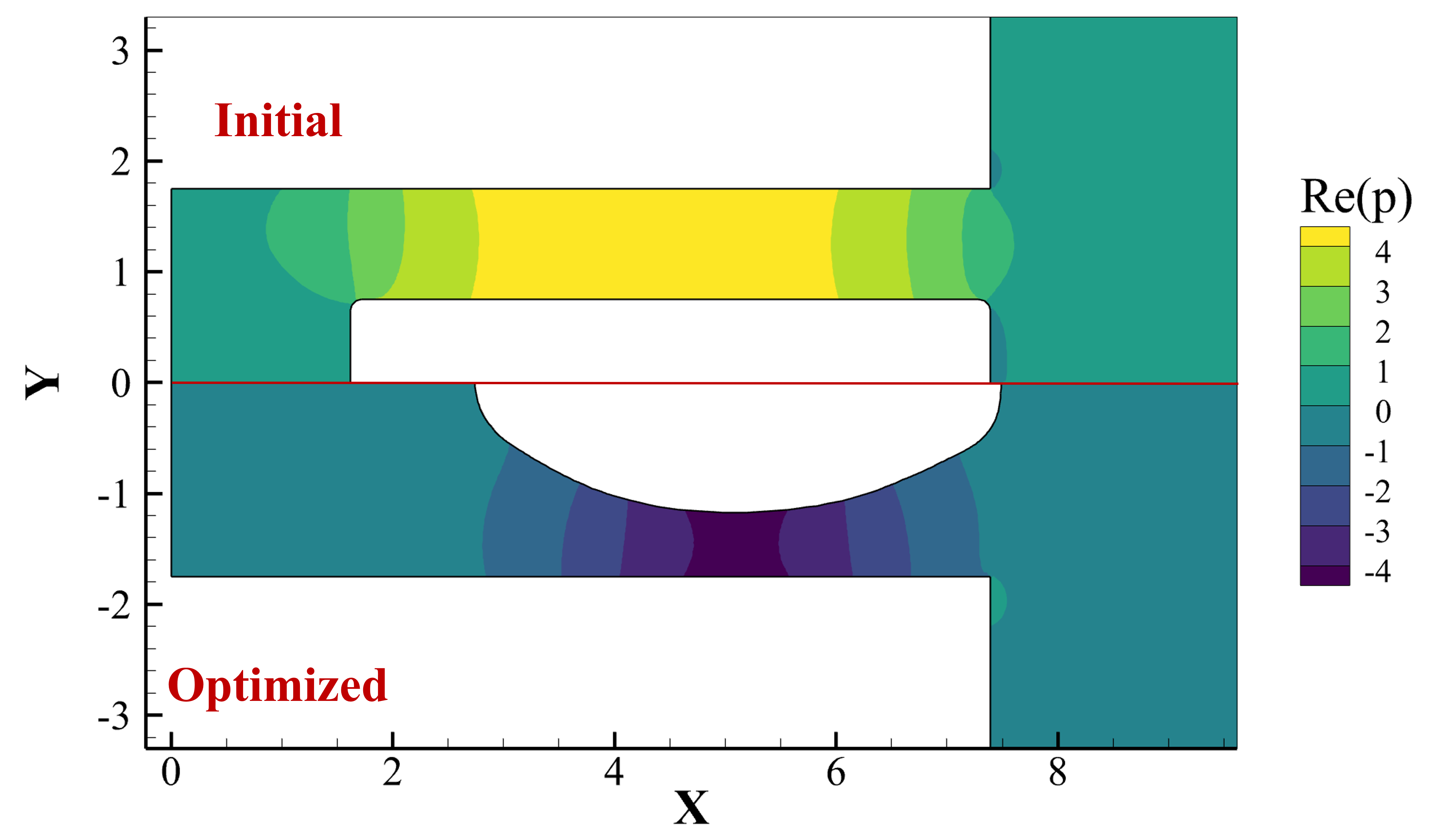}
    \hspace{0.1cm}
    \includegraphics[width=0.45\linewidth]{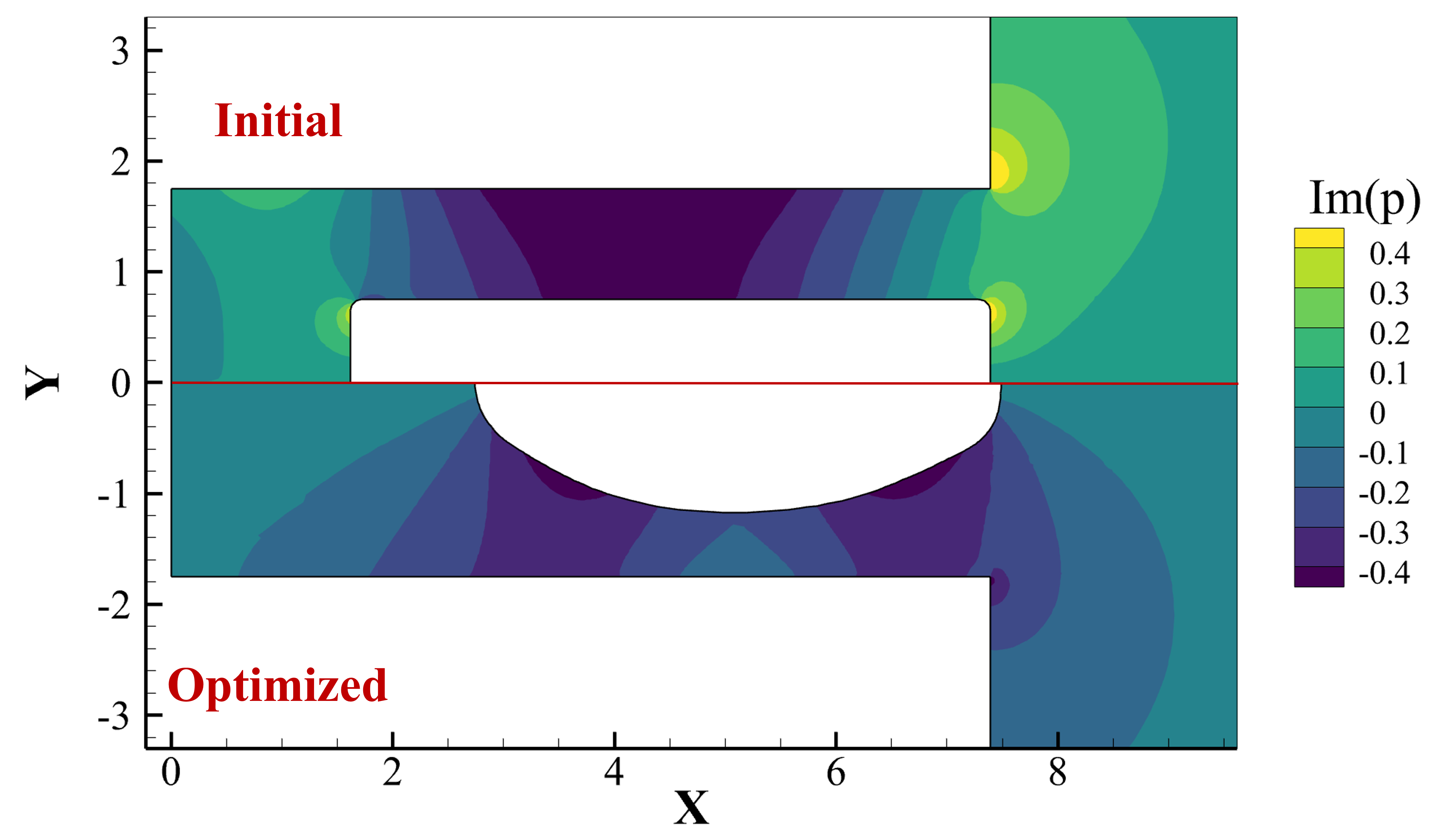}\\
    \vspace{0.3cm}
    \includegraphics[width=0.45\linewidth]{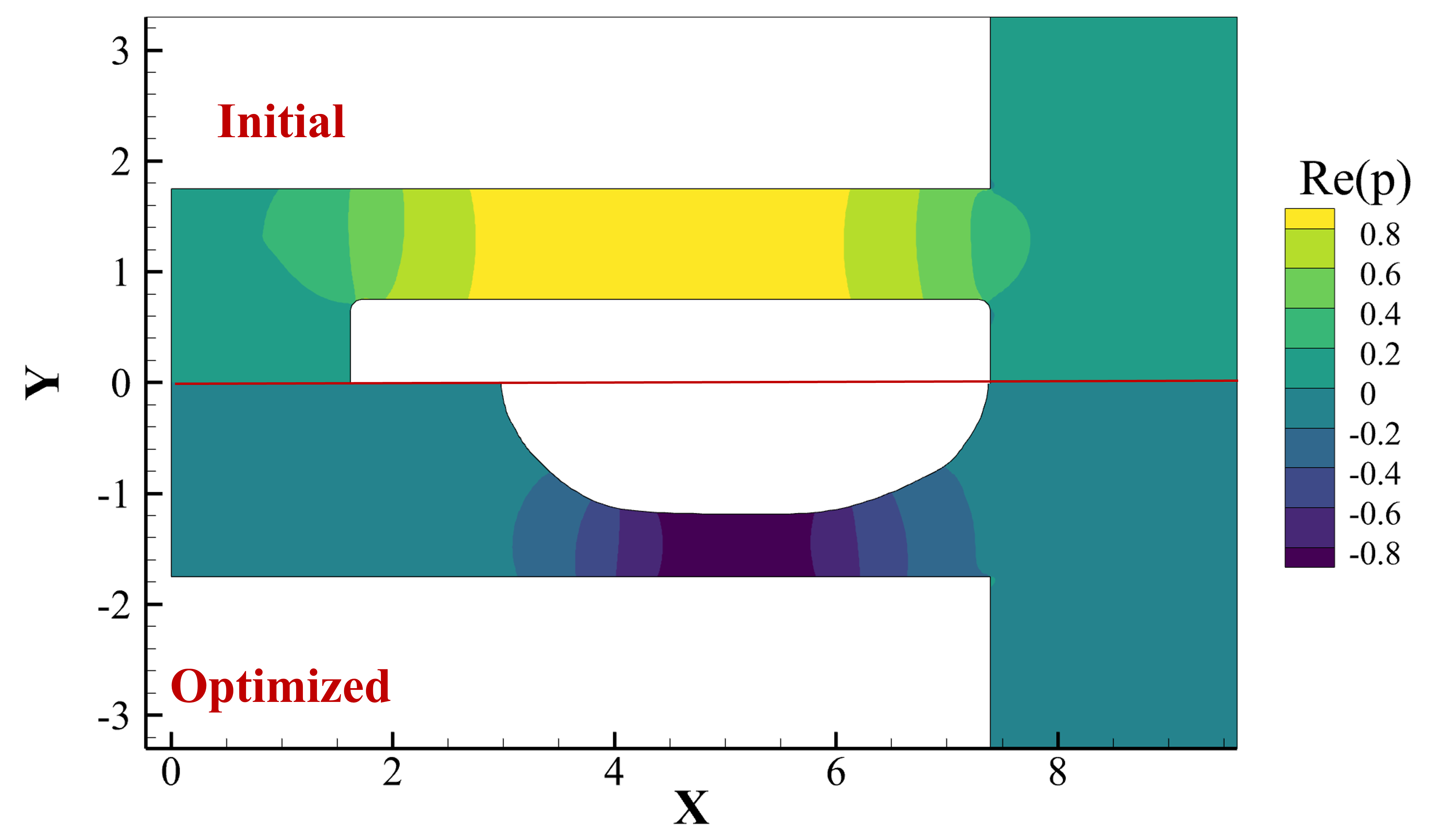}
    \hspace{0.1cm}
    \includegraphics[width=0.45\linewidth]{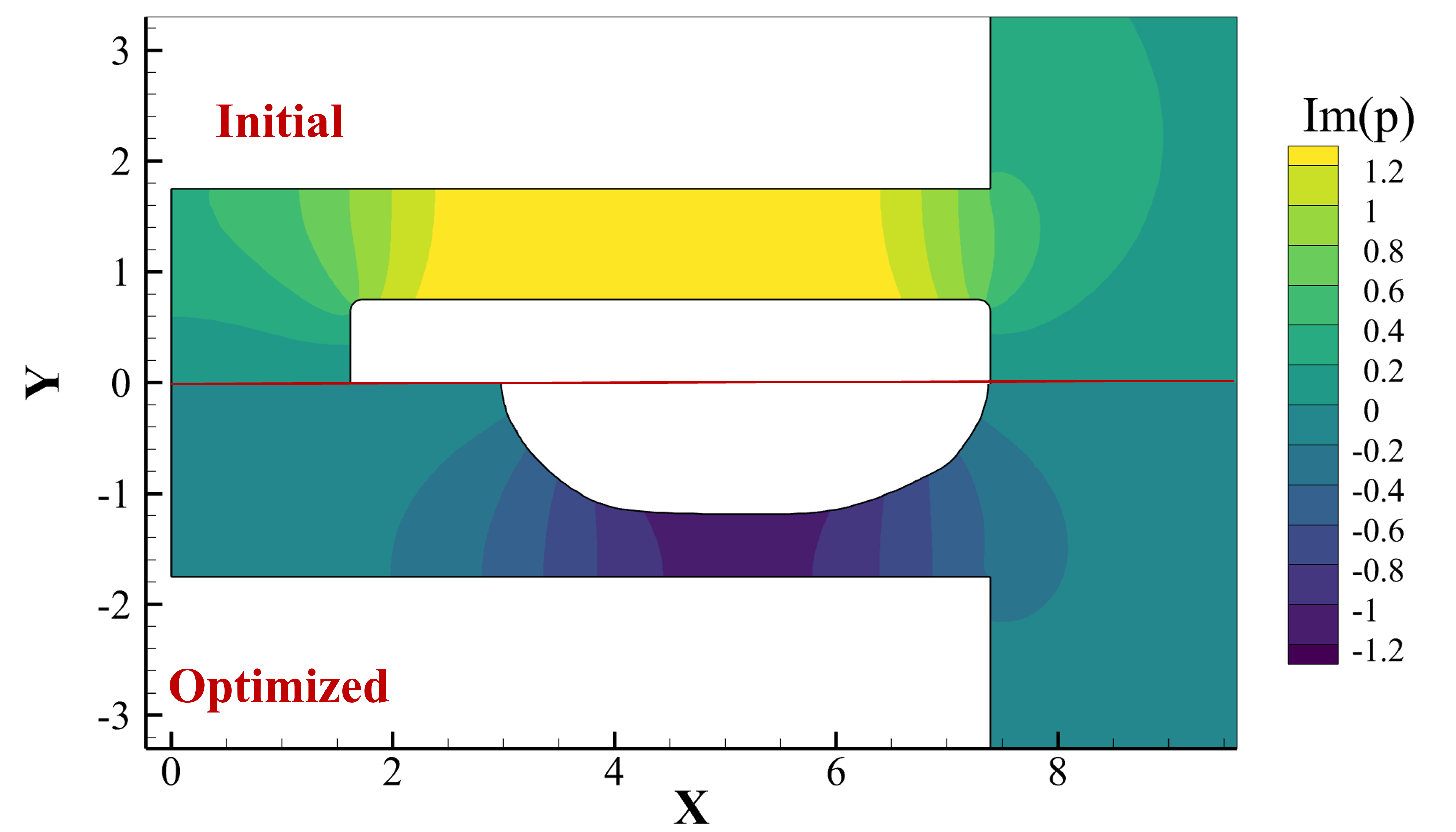}
    \caption{Comparison of the real and imaginary parts of the pressure perturbation around the initial and optimized geometries. The first, second, and third rows correspond to \((\delta_{rp},S)=(1,0.1)\), \((10,0.1)\), and \((100,0.1)\), respectively.}
    \label{fig:ba_pressure}
\end{figure}

Figure~\ref{fig:ba_pressure} compares the real and imaginary parts of the pressure perturbation around the rounded initial and optimized geometries. The optimized shape reflects the trade-off imposed by the area constraint. Because the vertical force is mainly generated by the pressure difference across the shuttle, the optimizer redistributes the shuttle area toward the right opening to facilitate pressure relief and reduces its horizontal extent to decrease the projected pressure-loaded area. Area constraint consequently increases the shuttle width and narrows the local gap, resulting in higher local pressure peaks. Nevertheless, the reduction in the pressure-loaded area outweighs the increase in local pressure, leading to a lower integrated vertical force.

The relative contributions of the complex force depend on the flow regime. For \((\delta_{rp},S)=(1,0.1)\), the reduction mainly arises from the imaginary part, whose magnitude decreases by approximately \(68\%\). For \((10,0.1)\), it is dominated by the approximately \(45\%\) decrease in the real part, while the imaginary part changes sign with little change in magnitude. For \((100,0.1)\), both parts are substantially reduced. Thus, the optimized geometry reduces the total force through different combinations of damping- and stiffness-related effects at different rarefaction levels.

We compare the computational costs of CIS and GSIS over the first ten optimization steps. The code is implemented in double precision with OpenMP parallelization and run on an AMD EPYC 7763 processor (2.45 GHz) using 8 threads. Under identical numerical and optimization settings, the wall‑clock times for CIS are \(0.50\), \(1.12\), and \(9.77\) hours for \(\delta_{rp}=1\), 10, and 100, respectively. The corresponding GSIS wall‑clock times are \(0.44\), \(0.27\), and \(0.23\) hours. The resulting speedup rises from \(1.13\) to \(42.5\), demonstrating that GSIS eliminates the severe convergence bottleneck of CIS in the near‑continuum regime.

\section{Conclusions}\label{sec:conclusion}

We develop an adjoint GSIS framework for shape optimization of linear oscillatory rarefied gas flows. We use the frequency-domain linearized Shakhov kinetic equation to model small-amplitude flow responses and derive a mesoscopic adjoint formulation to efficiently calculate gas damping force functionals. Macroscopic synthetic equations are built via velocity moments combined with Chapman–Enskog continuum closures and high-order rarefaction corrections. The adjoint GSIS offers fast convergence and asymptotic-preserving features for low-frequency near-continuum flows. Discrete shape derivatives are extracted from discretized boundary integrals to compute geometry gradients without repeated primal solves.

The proposed framework is validated through shape-optimization cases of an oscillating cylinder and a biaxial MEMS accelerometer. Comparisons with finite-difference and CIS results confirm the accuracy of the adjoint sensitivities, while the numerical solutions demonstrate the fast-converging and asymptotic-preserving properties of the adjoint GSIS. For the oscillating cylinder, the optimized geometries reduce the horizontal gas-force amplitude by \(33.1\%\), \(62.8\%\), and \(84.0\%\) for \((\delta_{rp},S)=(1,1)\), \((10,1)\), and \((100,1)\), respectively. For the biaxial accelerometer, whose baseline force agrees well with the experimental data, the optimized shuttle achieves corresponding reductions of \(35.5\%\), \(44.3\%\), and \(52.6\%\) under the prescribed geometric constraints. Moreover, GSIS avoids the rapid growth in computational cost encountered by CIS toward the near-continuum regime and delivers a speedup of \(42.5\) at \((\delta_{rp},S)=(100,0.1)\); the speedup is anticipated to increase as the rarefaction parameter rises.


Overall, this study creates a unified adjoint optimization tool for oscillatory rarefied gas flows. The fast-converging, asymptotic-preserving adjoint GSIS greatly reduces computational costs for gradient-based MEMS shape optimization, offering a reliable numerical framework for automated low-damping design of MEMS operated in rarefied gas environments.

\section*{Declaration of competing interest}
The authors declare that they have no known competing financial interests or personal relationships that could have appeared to influence the work reported in this paper.



\appendix

\section{Numerical schemes for the adjoint kinetic equation}

A cell-centered finite-volume method is employed to discretize the kinetic equations. The numerical discretization of the primal kinetic equation has been presented in Ref.~\cite{li2026frequency}. Therefore, only the adjoint GSIS is described here.

\subsection{Finite-volume discretization of the adjoint kinetic equation}\label{sec:finite_v_kinetic}

Let \(P\) denote a control volume and \(N\) its neighboring control volume sharing the face \(f\) with \(P\). The volume and centroid of cell \(P\) are denoted by \(V_P\) and \(\bm{x}_P\), respectively. The outward face-area vector with respect to cell \(P\) is defined as \(\bm{A}_f=A_f\bm{n}_f\), where \(A_f\) is the face area and \(\bm{n}_f\) is the unit normal pointing from \(P\) to \(N\). For a fixed discrete molecular velocity \(\bm{v}_k\), integrating Eq.~\eqref{eq:adjoint_shakhov} over cell \(P\) and applying the divergence theorem gives
\begin{equation}
    (\delta_{rp}+\mathrm{i}S)\phi_{P,k}^{n+\frac12}-\frac{1}{V_P}\sum_{f\in\partial P}\mathcal{F}_{f,k}^{n+\frac12}=\delta_{rp}\hat{\mathcal L}\phi_{P,k}^{n},
    \label{eq:fv_adjoint}
\end{equation}
where \(\partial P\) denotes the set of faces of cell \(P\), and \(n\) is the outer iteration index. For an internal face \(f\) shared by cells \(P\) and \(N\), the numerical flux is evaluated as
\begin{equation}
    \mathcal{F}_{f,k}^{n+\frac12}=A_f\left(\xi_{f,k}^{+}\phi_{f,k}^{N,n+\frac12}+\xi_{f,k}^{-}\phi_{f,k}^{P,n+\frac12}\right),\qquad \xi_{f,k}=\bm{v}_k\cdot\bm{n}_f,
    \label{eq:flux_def}
\end{equation}
where $\xi_{f,k}^{+}=\max(\xi_{f,k},0)$ and $\xi_{f,k}^{-}=\min(\xi_{f,k},0)$;
\(\phi_{f,k}^{P}\) and \(\phi_{f,k}^{N}\) are the face values reconstructed from cells \(P\) and \(N\), respectively. Since the transport operator in the adjoint equation is \(-\bm{v}\cdot\nabla\), the upwind direction is opposite to that of the primal kinetic equation. Therefore, the value reconstructed from cell \(N\) is selected when \(\xi_{f,k}>0\), whereas that reconstructed from cell \(P\) is selected when \(\xi_{f,k}<0\). 
A second-order linear reconstruction with explicitly evaluated gradients is adopted:
\begin{equation}
\begin{aligned}
    \phi_{f,k}^{P,n+\frac12}&=
    \phi_{P,k}^{n+\frac12}+\nabla\phi_{P,k}^{n}\cdot(\bm{x}_f-\bm{x}_P),\\
    \phi_{f,k}^{N,n+\frac12}&=
    \phi_{N,k}^{n+\frac12}+\nabla\phi_{N,k}^{n}\cdot(\bm{x}_f-\bm{x}_N),
\end{aligned}
\label{eq:reconstruction}
\end{equation}
where \(\bm{x}_f\) is the face centroid, and \(\bm{x}_N\) is the centroid of the neighboring cell \(N\). The cell-centered gradients are evaluated using a least-squares reconstruction based on neighboring cell values. For a boundary face, the exterior state is determined from the corresponding kinetic boundary condition.

The resulting discrete equation can be written as
\begin{equation}
    d_{P,k}\phi_{P,k}^{n+\frac12}+\sum_{N\in\mathcal N(P)}d_{PN,k}\phi_{N,k}^{n+\frac12}=b_{P,k}^{n},
    \label{eq:matrix_kinetic}
\end{equation}
where \(\mathcal N(P)\) denotes the set of cells neighboring \(P\), \(d_{P,k}\) and \(d_{PN,k}\) are the diagonal and off-diagonal coefficients generated by the implicit first-order upwind flux, and \(b_{P,k}^{n}\) contains the collision term, explicit reconstruction corrections, and boundary-flux contributions. The resulting sparse linear system is solved using the lower--upper symmetric Gauss--Seidel method~\cite{yoon1988lower}.

\subsection{Adjoint GSIS and incremental boundary treatment}
\label{sec:finite_v_gsis}

Within the CIS framework, the iteration described in \ref{sec:finite_v_kinetic} proceeds until convergence by setting $\phi^{n+\frac{1}{2}}$ equal to $\phi^{n+1}$. In the GSIS, at the $n$-th outer iteration, the finite-volume kinetic equation is first solved to obtain the intermediate adjoint distribution $\phi^{n+\frac{1}{2}}$. The corresponding macroscopic moments and high-order terms are then evaluated from $\phi^{n+\frac{1}{2}}$. Then, integrating Eqs.~\eqref{eq:adjoint_synthetic_macro_equations_rho} over a control volume $P$ gives
\begin{equation}
    \mathrm{i}SV_P
    \begin{bmatrix}
        2\hat{\rho}^{n+1}_P\\
        \hat{\bm u}^{n+1}_P\\
        \hat{\tau}^{n+1}_P
    \end{bmatrix}
    +
    \sum_{f\in\partial P}A_f
    \left(
        \widehat{\bm F}^{\mathrm{NS},n+1}_f
        -
        \widehat{\bm F}^{\mathrm{HoT},n+\frac{1}{2}}_f
    \right)
    =\bm 0,
    \label{eq:adjoint_synthetic_fvm}
\end{equation}
where for a face $f$ with outward unit normal $\bm n_f$, the NS and high-order fluxes per unit area are defined as
\begin{equation}
    \widehat{\bm F}^{\mathrm{NS}}_f
    =
    \begin{bmatrix}
        -\hat{\bm u}_f\cdot\bm n_f\\
        -\left(\hat{\rho}_f+\dfrac{2}{3}\hat{\tau}_f\right)\bm n_f-\widehat{\bm{\Pi}}^{\mathrm{NS}}_f\cdot\bm n_f\\
        -\dfrac{1}{2}\hat{\bm u}_f\cdot\bm n_f-\dfrac{15}{4}\widehat{\bm q}^{\mathrm{NS}}_f\cdot\bm n_f
    \end{bmatrix},
    \qquad
    \widehat{\bm F}^{\mathrm{HoT}}_f
    =
    \begin{bmatrix}
        0\\
        \widehat{\bm{\Pi}}^{\mathrm{HoT}}_f\cdot\bm n_f\\
        \dfrac{15}{4}\widehat{\bm q}^{\mathrm{HoT}}_f\cdot\bm n_f
    \end{bmatrix}.
    \label{eq:adjoint_macro_flux}
\end{equation}
The face interpolation, Rhie--Chow-type correction, and fully implicit block-coupled solution follow Refs.~\cite{li2026frequency,darwish2009coupled}.

At an internal face, the fluxes in Eq.~\eqref{eq:adjoint_synthetic_fvm} are evaluated from the adjacent macroscopic states. At a physical boundary, however, fixing the boundary flux at its kinetic value prevents it from responding to the evolving synthetic solution. Following Zhang \textit{et al.}~\cite{zhang2026increment}, we introduce an incremental half-space treatment for the adjoint kinetic system.


Define the adjoint moment vector
\begin{equation}
    \widehat{\bm\eta}(\bm v)
    =\left[2,2\bm v,|\bm v|^2-\dfrac{3}{2}\right]^\top.
    \label{eq:adjoint_boundary_moment_vector}
\end{equation}
For a boundary face $b$ with outward unit normal $\bm n_b$, the reference flux is evaluated from the intermediate kinetic distribution as
\begin{equation}
    \widehat{\bm F}_b^0
    =
    -\int_{\Xi}
    v_n\widehat{\bm\eta}(\bm v)
    \phi_b^{n+\frac{1}{2}}(\bm v)
    f_{\mathrm{eq}}(\bm v)\,\mathrm{d}\Xi,
    \qquad
    v_n=\bm v\cdot\bm n_b.
    \label{eq:adjoint_boundary_reference_flux}
\end{equation}
During the inner iteration of macroscopic synthetic equation \eqref{eq:adjoint_synthetic_fvm}, the boundary flux is updated according to
\begin{equation}
    \widehat{\bm F}_b^m
    =
    \widehat{\bm F}_b^0
    +
    \Delta\widehat{\bm F}_b^m,
    \qquad
    \Delta\widehat{\bm U}_b^m
    =
    \widehat{\bm U}_b^m-\widehat{\bm U}_b^0,
    \qquad
    \widehat{\bm U}
    =
    \begin{bmatrix}
        \hat{\rho} & \hat{\bm u} & \hat{\tau}
    \end{bmatrix}^\top,
    \label{eq:adjoint_boundary_flux_update}
\end{equation}
where $m$ denotes the inner-iteration index, $\widehat{\bm U}_b^0$ is evaluated from $\phi_b^{n+\frac{1}{2}}$, and $\Delta\widehat{\bm F}_b^m$ is the linearized flux correction induced by $\Delta\widehat{\bm U}_b^m$. At a boundary face, the flux $\widehat{\bm F}^{\mathrm{NS}}_f-\widehat{\bm F}^{\mathrm{HoT}}_f$ in Eq.~\eqref{eq:adjoint_synthetic_fvm} is replaced by $\widehat{\bm F}_b^m$.

The corresponding flux increment is
\begin{equation}
    \Delta\widehat{\bm F}_b
    =
    -\Delta\widehat{\bm H}_b^+
    -
    \Delta\widehat{\bm H}_b^-,
    \qquad
    \Delta\widehat{\bm H}_b^\pm
    =
    \int_{\Xi^\pm}
    v_n\widehat{\bm\eta}(\bm v)
    \Delta\phi^\pm
    f_{\mathrm{eq}}(\bm v)\,\mathrm{d}\Xi.
    \label{eq:adjoint_boundary_half_fluxes}
\end{equation}
Because the transport operator in the adjoint equation is $-\bm v\cdot\nabla$, the distribution on $\Xi^-$ is determined by the interior solution. Its increment is approximated by the local adjoint equilibrium increment,
\begin{equation}
    \Delta\phi^-
    \approx
    \hat{\mathcal L}(\Delta\phi)
    =
    \Delta\hat{\rho}
    +
    \bm v\cdot\Delta\hat{\bm u}
    +
    \left(
        \frac{2}{3}|\bm v|^2-1
    \right)
    \Delta\hat{\tau},
    \qquad
    v_n<0.
    \label{eq:adjoint_equilibrium_increment}
\end{equation}
Introducing the normal and tangential velocity increments
\begin{equation}
    \Delta\hat u_n
    =
    \Delta\hat{\bm u}\cdot\bm n_b,
    \qquad
    \Delta\hat{\bm u}_t
    =
    \Delta\hat{\bm u}
    -
    \Delta\hat u_n\bm n_b,
    \label{eq:adjoint_velocity_increment_decomposition}
\end{equation}
the half-space moments of Eq.~\eqref{eq:adjoint_equilibrium_increment} give
\begin{equation}
    \Delta\widehat{\bm H}_b^-
    =
    \begin{bmatrix}
        -\dfrac{\Delta\hat{\rho}}{\sqrt{\pi}}
        +\dfrac{1}{2}\Delta\hat u_n
        -\dfrac{\Delta\hat{\tau}}{3\sqrt{\pi}}\\[6pt]
        \left(
            \dfrac{1}{2}\Delta\hat{\rho}
            -\dfrac{\Delta\hat u_n}{\sqrt{\pi}}
            +\dfrac{1}{3}\Delta\hat{\tau}
        \right)\bm n_b
        -
        \dfrac{\Delta\hat{\bm u}_t}{2\sqrt{\pi}}\\[6pt]
        -\dfrac{\Delta\hat{\rho}}{4\sqrt{\pi}}
        +\dfrac{1}{4}\Delta\hat u_n
        -\dfrac{3\Delta\hat{\tau}}{4\sqrt{\pi}}
    \end{bmatrix}.
    \label{eq:adjoint_interior_half_flux_increment}
\end{equation}

For a diffuse-reflection wall, the adjoint boundary condition determines the distribution increment on $\Xi^+$ from that on $\Xi^-$:
\begin{equation}
    \Delta\phi^+
    =
    -2\sqrt{\pi}
    \int_{\Xi^-}
    v_n\Delta\phi^-f_{\mathrm{eq}}\,\mathrm{d}\Xi
    \approx
    \Delta\hat{\rho}
    -
    \frac{\sqrt{\pi}}{2}\Delta\hat u_n
    +
    \frac{1}{3}\Delta\hat{\tau},
    \qquad
    v_n>0.
    \label{eq:adjoint_diffuse_distribution_increment}
\end{equation}
Substituting Eqs.~\eqref{eq:adjoint_interior_half_flux_increment} and \eqref{eq:adjoint_diffuse_distribution_increment} into Eq.~\eqref{eq:adjoint_boundary_half_fluxes} gives the diffuse-wall flux correction
\begin{equation}
    \Delta\widehat{\bm F}_b^{\mathrm W}
    =
    \begin{bmatrix}
        0\\[2pt]
        \left[
            -\Delta\hat{\rho}
            +
            \left(
                \dfrac{1}{\sqrt{\pi}}
                +
                \dfrac{\sqrt{\pi}}{4}
            \right)
            \Delta\hat u_n
            -
            \dfrac{1}{2}\Delta\hat{\tau}
        \right]\bm n_b
        +
        \dfrac{\Delta\hat{\bm u}_t}{2\sqrt{\pi}}\\[6pt]
        -\dfrac{1}{8}\Delta\hat u_n
        +
        \dfrac{2}{3\sqrt{\pi}}\Delta\hat{\tau}
    \end{bmatrix}.
    \label{eq:adjoint_diffuse_flux_increment}
\end{equation}

At a Dirichlet boundary, the prescribed distribution on $\Xi^+$ remains fixed during the macroscopic inner iterations, and hence \(\Delta\phi^+=0\). Consequently, no $\Xi^+$ correction is introduced, and Eq.~\eqref{eq:adjoint_boundary_half_fluxes} reduces to
\begin{equation}
    \Delta\widehat{\bm F}_b^{\mathrm D}
    =
    -\Delta\widehat{\bm H}_b^-.
\end{equation}
Using Eq.~\eqref{eq:adjoint_interior_half_flux_increment}, the explicit Dirichlet boundary correction is
\begin{equation}
    \Delta\widehat{\bm F}_b^{\mathrm D}
    =
    \begin{bmatrix}
        \dfrac{\Delta\hat{\rho}}{\sqrt{\pi}}
        -\dfrac{1}{2}\Delta\hat u_n
        +\dfrac{\Delta\hat{\tau}}{3\sqrt{\pi}}\\[6pt]
        \left(
            -\dfrac{1}{2}\Delta\hat{\rho}
            +\dfrac{\Delta\hat u_n}{\sqrt{\pi}}
            -\dfrac{1}{3}\Delta\hat{\tau}
        \right)\bm n_b
        +
        \dfrac{\Delta\hat{\bm u}_t}{2\sqrt{\pi}}\\[6pt]
        \dfrac{\Delta\hat{\rho}}{4\sqrt{\pi}}
        -\dfrac{1}{4}\Delta\hat u_n
        +\dfrac{3\Delta\hat{\tau}}{4\sqrt{\pi}}
    \end{bmatrix}.
    \label{eq:adjoint_dirichlet_flux_increment}
\end{equation}

\subsection{Two-dimensional reduced formulation}
\label{app:two_dimensional_reduced_formulation}

For the two-dimensional configurations considered in this work, the physical fields are independent of the $z$ coordinate, whereas the molecular velocity remains three-dimensional. Let $\bm c=(v_x,v_y)$ denote the in-plane molecular velocity and $c^2=|\bm c|^2$. To avoid discretizing the out-of-plane velocity $v_z$, its dependence is integrated analytically by introducing
\begin{equation}\label{eq:reduced_distribution_definition}
\begin{aligned}
    h_1(\bm x,\bm c)
    =&
    \frac{1}{\sqrt{\pi}}
    \int_{-\infty}^{\infty}
    h(\bm x,\bm c,v_z)e^{-v_z^2}\,\mathrm{d}v_z,
    \\
    h_2(\bm x,\bm c)
    =&
    \frac{2}{\sqrt{\pi}}
    \int_{-\infty}^{\infty}
    v_z^2h(\bm x,\bm c,v_z)e^{-v_z^2}\,\mathrm{d}v_z,  
\end{aligned}
\end{equation}
where $\bm x=(x,y)$. The second reduced distribution retains the contribution of the out-of-plane molecular energy. The two-dimensional velocity average is denoted by
\begin{equation}
    \langle a\rangle
    =
    \int_{\Xi}
    a(\bm c)f_{\mathrm{eq}}^{2\mathrm D}(\bm c)\,\mathrm{d}\bm c,
    \qquad
    f_{\mathrm{eq}}^{2\mathrm D}(\bm c)
    =
    \frac{1}{\pi}e^{-c^2}.
    \label{eq:reduced_velocity_average}
\end{equation}

The reduced adjoint formulation is constructed from the reduced primal equation, objective functional, and boundary conditions rather than by directly integrating the three-dimensional adjoint equation over $v_z$. Since the corresponding Lagrangian variation follows the same procedure as the three-dimensional derivation, only the resulting reduced equations are given below.

The reduced primal kinetic equation is
\begin{equation}
    \mathrm{i}S\bm h+\bm c\cdot\nabla\bm h
    =
    \delta_{\mathrm{rp}}(\bm g-\bm h),
    \qquad
    \bm h=
    \begin{bmatrix}
        h_1\\
        h_2
    \end{bmatrix},
    \qquad
    \bm g=
    \begin{bmatrix}
        g_1\\
        g_2
    \end{bmatrix},
    \label{eq:reduced_forward_kinetic_equation}
\end{equation}
where $\nabla=(\partial_x,\partial_y)$ and
\begin{equation}
    \begin{aligned}
        g_1
        &=
        \rho+2\bm c\cdot\bm u+(c^2-1)\tau
        +\frac{4}{15}(c^2-2)\bm c\cdot\bm q,\\
        g_2
        &=
        \rho+2\bm c\cdot\bm u+c^2\tau
        +\frac{4}{15}(c^2-1)\bm c\cdot\bm q.
    \end{aligned}
    \label{eq:reduced_reference_distribution}
\end{equation}
The macroscopic variables are recovered from
\begin{equation}
    \begin{aligned}
        \rho
        &=
        \langle h_1\rangle,
        &
        \bm u
        &=
        \langle\bm c h_1\rangle,\\
        \tau
        &=
        \left\langle
        \left(\frac{2}{3}c^2-1\right)h_1+\frac{1}{3}h_2
        \right\rangle,
        &
        \bm q
        &=
        \left\langle
        \left(c^2-\frac{5}{2}\right)\bm c h_1
        +\frac{1}{2}\bm c h_2
        \right\rangle.
    \end{aligned}
    \label{eq:reduced_forward_moments}
\end{equation}

Let $c_n=\bm c\cdot\bm n$ and define the two-dimensional half velocity spaces as
\begin{equation}
    \Xi^+=\{\bm c\in\Xi:c_n>0\},
    \qquad
    \Xi^-=\{\bm c\in\Xi:c_n<0\}.
    \label{eq:reduced_half_velocity_spaces}
\end{equation}
For a diffuse-reflection wall, the reduced boundary condition is
\begin{equation}
    h_1(\bm c)
    =
    h_2(\bm c)
    =
    2\sqrt{\pi}
    \int_{\Xi^+}
    (\bm c'\cdot\bm n)h_1(\bm c')
    f_{\mathrm{eq}}^{2\mathrm D}(\bm c')\,\mathrm{d}\bm c'
    +
    2\bm u_w\cdot\bm c
    -
    \sqrt{\pi}\bm u_w\cdot\bm n,
    \qquad
    \bm c\in\Xi^-.
    \label{eq:reduced_forward_wall_condition}
\end{equation}
At the Dirichlet boundary, we have $\bm h=\bm h_d$ for $ \bm c\in\Xi^-$,
where $\bm h_d$ is obtained by reducing the prescribed three-dimensional boundary distribution. For the equilibrium far-field condition considered here, $\bm h_d=\bm 0$.

For the in-plane force objectives considered in this work, the moment kernel is independent of $v_z$, and the reduced objective therefore depends only on $h_1$:
\begin{equation}
    J
    =
    \int_{\Gamma_m}
    \int_{\Xi}
    c_nm(\bm c)h_1
    f_{\mathrm{eq}}^{2\mathrm D}(\bm c)\,\mathrm{d}\bm c\,\mathrm{d}\Gamma.
    \label{eq:reduced_boundary_response}
\end{equation}

Applying the same Lagrangian variation as in the three-dimensional formulation to the reduced primal system gives
\begin{equation}
    \mathrm{i}S\bm\phi-\bm c\cdot\nabla\bm\phi
    =
    \delta_{\mathrm{rp}}
    \left(
        \bm\phi_{\mathrm{eq}}-\bm\phi
    \right),
    \qquad
    \bm\phi=
    \begin{bmatrix}
        \phi_1\\
        \phi_2
    \end{bmatrix}.
    \label{eq:reduced_adjoint_kinetic_equation}
\end{equation}
The reduced gain term in the adjoint collision operator is
\begin{equation}
    \begin{aligned}
        \phi_{\mathrm{eq},1}
        &=
        \hat{\rho}
        +
        \bm c\cdot\hat{\bm u}
        +
        \left(
            \frac{2}{3}c^2-1
        \right)\hat{\tau}
        +
        \left(
            c^2-\frac{5}{2}
        \right)\bm c\cdot\hat{\bm q},\\
        \phi_{\mathrm{eq},2}
        &=
        \frac{1}{3}\hat{\tau}
        +
        \frac{1}{2}\bm c\cdot\hat{\bm q},
    \end{aligned}
    \label{eq:reduced_adjoint_reference_distribution}
\end{equation}
where the corresponding adjoint moments are
\begin{equation}
    \begin{aligned}
        \hat{\rho}
        &=
        \langle\phi_1+\phi_2\rangle,
        &
        \hat{\bm u}
        &=
        \langle2\bm c(\phi_1+\phi_2)\rangle,\\
        \hat{\tau}
        &=
        \left\langle
        (c^2-1)\phi_1+c^2\phi_2
        \right\rangle,
        &
        \hat{\bm q}
        &=
        \frac{4}{15}
        \left\langle
        (c^2-2)\bm c\phi_1
        +(c^2-1)\bm c\phi_2
        \right\rangle.
    \end{aligned}
    \label{eq:reduced_adjoint_moments}
\end{equation}


The reduced adjoint stress and higher-order moments used in the macroscopic synthetic equation are then evaluated as
\begin{equation}
    \begin{aligned}
        \hat{\Pi}_{ij}
        &=
        2\left\langle
        \mathcal A_{ij}^{(1)}\phi_1
        +
        \mathcal A_{ij}^{(2)}\phi_2
        \right\rangle,\\
        \hat{M}_{ijk}
        &=
        2\left\langle
        \mathcal A_{ij}^{(1)}c_k\phi_1
        +
        \mathcal A_{ij}^{(2)}c_k\phi_2
        \right\rangle,\\
        \hat{Q}_{ij}
        &=
        \frac{4}{15}
        \left\langle
        (c^2-2)c_ic_j\phi_1
        +
        (c^2-1)c_ic_j\phi_2
        \right\rangle,
    \end{aligned}
    \label{eq:reduced_adjoint_high_order_moments}
\end{equation}
where $\mathcal A_{ij}^{(1)}
    =
    c_ic_j
    -
    \frac{1}{3}
    \left(
        c^2+\frac{1}{2}
    \right)\delta_{ij}$ and 
    $\mathcal A_{ij}^{(2)}
    =
    c_ic_j
    -
    \frac{1}{3}
    \left(
        c^2+\frac{3}{2}
    \right)\delta_{ij}$.
These reduced moments are substituted directly into Eq.~\eqref{eq:adjoint_hot_definition} to evaluate $\widehat{\bm\Pi}^{\mathrm{HoT}}$ and $\widehat{\bm q}^{\mathrm{HoT}}$.

For a diffuse-reflection wall, the reduced adjoint boundary condition is
\begin{equation}
    \begin{aligned}
        \phi_1 &=
        -m(\bm c)
        -
        2\sqrt{\pi}
        \int_{\Xi^-}
        (\bm{c}' \cdot \bm{n})
        \left[
            \phi_1
            +
            \phi_2
            +
            m(\bm c')
        \right]
        f_{\mathrm{eq}}^{2\mathrm D}(\bm c')\,\mathrm{d}\bm c',\\
        \phi_2 &= 0,
    \end{aligned}
    \qquad
    \bm c\in\Xi^+.
    \label{eq:reduced_adjoint_wall_condition}
\end{equation}

At the Dirichlet boundary, we have $\bm\phi=\bm 0$ for $\bm c\in\Xi^+$.

\bibliographystyle{elsarticle-num}

\bibliography{cite}

\end{CJK}
\end{document}